\documentclass{article}

\usepackage{amsthm}
\usepackage{amsmath}
\usepackage{amssymb}
\usepackage{bm}
\usepackage{hyperref}
\usepackage[noabbrev,capitalise,nosort,nameinlink]{cleveref}
\usepackage{natbib}
\usepackage{mathtools}
\usepackage{lineno}
\usepackage{multirow}
\usepackage{caption}
\usepackage{subcaption}

\newtheorem{theorem}{\sffamily Theorem}
\DeclareMathOperator{\Tr}{Tr}

\newcommand{\todo}[1]{\bgroup\color{red}#1\egroup}

\usepackage[a4paper, top=88pt, bottom=88pt, left=72pt, right=72pt, headsep=16pt, footskip=28pt]{geometry}
\hypersetup{
	colorlinks,
	linkcolor=blue,
	citecolor=blue,
	urlcolor=blue,
}
\newcommand*{\email}[1]{\bgroup\color{blue}\href{mailto:#1}{#1}\egroup}
\renewcommand*{\url}[1]{\bgroup\color{blue}\href{#1}{#1}\egroup}
\usepackage{enumitem}
\setlist[enumerate]{nosep}
\setlist[itemize]{nosep}
\usepackage{mleftright} \mleftright
\renewcommand{\qedsymbol}{$\blacksquare$}
\renewenvironment{proof}[1][\proofname]{\noindent{\sffamily\bfseries #1.} }{\hfill \qedsymbol \medskip}
\usepackage{scrlayer-scrpage, xhfill}
\usepackage[labelfont={sf,bf}]{caption}
\automark[section]{section}
\setkomafont{pageheadfoot}{\normalcolor\sffamily}
\setkomafont{pagenumber}{\normalfont\normalsize\sffamily}
\clearpairofpagestyles
\let\oldtitle\title
\renewcommand{\title}[1]{\oldtitle{#1}\newcommand{\theshorttitle}{#1}}

\let\oldauthor\author
\renewcommand{\author}[1]{\oldauthor{#1}\newcommand{\theshortauthor}{#1}}
\newcommand{\shortauthor}[1]{\renewcommand{\theshortauthor}{#1}}
\cohead{\xrfill[0.525ex]{0.6pt}~\theshorttitle~\xrfill[0.525ex]{0.6pt}}
\cehead{\xrfill[0.525ex]{0.6pt}~\theshortauthor~\xrfill[0.525ex]{0.6pt}}
\newcommand{\theabstract}[1]{\par\bgroup\noindent\textbf{\textsf{Abstract.}} #1\egroup}
\newcommand{\thekeywords}[1]{\par\smallskip\bgroup\noindent\textbf{\textsf{Keywords.}}\newcommand{\and}{ $\bullet$ } #1\egroup}
\newcommand{\themsc}[1]{\par\smallskip\bgroup\noindent\textbf{\textsf{2010 Mathematics Subject Classification.}}\newcommand{\and}{ $\bullet$ } #1\egroup}
\newcommand*{\affilref}[1]{\ref{affiliation#1}}
\newcommand*{\affiliation}[3]{
	\footnotetext[#1]{\label{affiliation#2} #3}
}

\title{Bayesian material flow analysis for systems with multiple levels of disaggregation and high dimensional data}

\author{%
	Junyang Wang\textsuperscript{\affilref{Imperial},\affilref{Imperial3}}
	\and
	Kolyan Ray\textsuperscript{\affilref{Imperial3}}
	\and
	Pablo Brito-Parada\textsuperscript{\affilref{Imperial2}}
	\and
	Yves Plancherel\textsuperscript{\affilref{Imperial2}}
	\and 
	Tom Bide\textsuperscript{\affilref{BGS}}
	\and 
	Joseph Mankelow\textsuperscript{\affilref{BGS}}
	\and 
	John Morley\textsuperscript{\affilref{Imperial2}}
	\and
	Julia Stegemann\textsuperscript{\affilref{UCL}}
	\and
	Rupert Myers\textsuperscript{\affilref{Imperial}}
	}
\shortauthor{J.~Wang, Rupert Myers}

\begin{document}

\maketitle
\affiliation{1}{Imperial}{Department of Civil and Environmental Engineering, Imperial College London, United Kingdom. \email{junyang.wang21@imperial.ac.uk}, \email{r.myers@imperial.ac.uk}}
\affiliation{2}{Imperial2}{Department of Earth Science and Engineering, Imperial College London, United Kingdom. \email{p.brito-parada@imperial.ac.uk},\email{y.plancherel@imperial.ac.uk}, \email{john.morley18@imperial.ac.uk}}
\affiliation{3}{Imperial3}{Department of Mathematics, Imperial College London, 
London, United Kingdom. \email{kolyan.ray@imperial.ac.uk}}
\affiliation{4}{UCL}{Department of Civil, Environmental and Geomatic Engineering, University College London, London, United Kingdom. \email{j.stegemann@ucl.ac.uk}}
\affiliation{5}{BGS}{British Geological Survey, United Kingdom. \email{tode@bgs.ac.uk}, \email{jmank@bgs.ac.uk}}

\begin{abstract}
	\theabstract{Material Flow Analysis (MFA) is used to quantify and understand the life cycles of materials from production to end of use, which enables environmental, social and economic impacts and interventions. MFA is challenging as available data is often limited and uncertain, leading to an underdetermined system with an infinite number of possible stocks and flows values. Bayesian statistics is an effective way to address these challenges by principally incorporating domain knowledge, and quantifying uncertainty in the data and providing probabilities associated with model solutions.

This paper presents a novel MFA methodology under the Bayesian framework. By relaxing the mass balance constraints, we improve the computational scalability and reliability of the posterior samples compared to existing Bayesian MFA methods. We propose a mass based, child and parent process framework to model systems with disaggregated processes and flows. We show posterior predictive checks can be used to identify inconsistencies in the data and aid noise and hyperparameter selection. The proposed approach is demonstrated on case studies, including a global aluminium cycle with significant disaggregation, under weakly informative priors and significant data gaps to investigate the feasibility of Bayesian MFA. We illustrate just a weakly informative prior can greatly improve the performance of Bayesian methods, for both estimation accuracy and uncertainty quantification.
\nolinebreak}
	
	\thekeywords{{Material flow analysis}\and{Bayesian statistics}\and{Probabilistic modelling}\and{Uncertainty quantification}\and{Missing data}\and{Circular Economy}}

\end{abstract}

\section{Introduction} \label{sec: introduction}

Increasingly, the world is facing major shifts in resource utilisation. To prevent the worst consequences of climate change, global carbon emissions must be significantly reduced, which requires a fundamental change in how fossil fuels and other high carbon footprint materials are used in the coming decades. Additionally, the instability of supply chains, changing behaviour caused by Covid-19 (\cite{Jowitt2020}), as well as the growing world population, projected to reach almost 10 billion by 2050 (\cite{WPP}), will also likely significantly impact resource utilisation globally (\cite{Jowitt2020nature}). However, many material flows of interest are poorly understood, which severely impedes the ability of policy makers to identify ways to use important materials or resources more efficiently and sustainably, and plan the transition to more sustainable systems of production and use (\cite{Mudd2021,UNECE2020}). 

Material flow analysis (MFA) is a broad terminology encompassing any quantitative method used to map and quantify flows and stocks of a material of interest in a well defined system. MFA is applicable to all materials and a range of economic, environmental and social scales, from a company's supply chain to entire countries and economies. MFA studies have been conducted on metals, including steel and aluminium (\cite{Bertram2009} and \cite{Cullen2012}), non-metals such as glass and concrete (\cite{Westbroek2021} and \cite{daCostaReis2019}), for entire countries (\cite{Matsubae2009}), on a global scale (\cite{Miatto2017}) and for multiple materials across whole economies (\cite{Fischer2011}. Common to all MFA models is the precise definition of a system, which includes the system boundary, the processes within this system, as well as stock and flow variables representing the storage within and movement between the processes. A physical assumption common to all MFA models is the conservation of mass, where the outflows and inflows of a process must balance with its change of stock. 

Typically in static MFA problems it is difficult or impossible to collect data for all the flow and change in stock variables inside the system. This gives rise to an underdetermined system, in the sense that the size of available data combined with physical constraints, containing at least mass balance of processes, is still less than the number of model parameters. For modelling this places MFA datasets in the domain of high dimensional statistics (\cite{Lederer2022}), which is statistically challenging as the number of unknown parameters exceeds the sample size (\cite{Wainwright2019}). Well known statistical methods in this area include ridge regression, the LASSO, Elastic net, as well as Bayesian approaches (\cite{Hoerl2000,Tibshirani1996,Zou2005,Kruschke2015}).

Bayesian statistics is particularly well suited for MFA for three main reasons. Firstly, Bayesian statistics works naturally in the underdetermined setting by providing a posterior probability over all possible solutions instead of a single solution. Secondly, while commonly there is a lack of data in MFA problems, there is often domain knowledge or expert opinion that can be used to greatly improve the prediction accuracy of the model, and the Bayesian prior distribution provides a natural framework to incorporate such domain knowledge, which is missed in the traditional MFA approach. Thirdly, Bayesian statistics rigorously quantifies uncertainty in the data and propagates it into the posterior distribution. Additionally, authors including \cite{Brunner2004} and \cite{Lupton2018} argued for an incremental approach to MFA, where the system diagram and data is continuously refined and improved until the required level of data certainty has been achieved. The Bayesian paradigm naturally faciliates this through iterative learning of models as new data becomes available by rerunning the model with new data, since the posterior distribution from a previous analysis can be interpreted as the prior distribution for the subsequent analysis.

Non Bayesian approaches to uncertainty quantification in MFA include works such as \cite{Laner2015}, \cite{Schwab2016}, \cite{Schwab2017}. These approaches focus on qualitative and quantitative methods of assigning confidence onto data or mass balanced flows, but do not provide a modelling procedure that combines prior belief with data and mass balance to produce flow estimates, nor a framework for propagating uncertainty. Bayesian inference provides a modelling procedure that is able to propagate uncertainty, and a rigorous, probabilistic interpretation of uncertainty. The popular software `STAN' (\cite{Cencic2016}) uses least squares minimisation to conduct data reconciliation of MFA systems. However, STAN requires nonlinear data to be approximated by first order Taylor expansion, and a normality assumption is made on flows, which can permit negative flow values that are not physically meaningful.

\subsection{Previous work on Bayesian approaches to MFA} \label{subsec: previousBayesianMFA}

The use of Bayesian statistics in MFA is still relatively limited. Perhaps the earliest use of Bayesian statistics in MFA is by \cite{Gottschalk2010},  where the authors proposed a model where the mass balance equations are parametrised using specific flow ratios (known as transfer coefficients) in the system, and tested the model on a case study of flows of concentrations of nanoparticles in Switzerland using simulated data. In \cite{Lupton2018}, the authors used the same mass balance parametrisation as \cite{Gottschalk2010}, but with Dirichlet priors on the transfer coefficients rather than uniform or triangular priors, using a HMC (Hamiltonian Monte Carlo) based sampler called the NUTS (No-U-Turn Sampler) algorithm to sample from the posterior, and conducted a case study mapping global steel flows. Recently, \cite{Dong2022} proposed a method which combines prior information from multiple experts in the MFA setting, demonstrating it using the mass balance parametrisation of \cite{Gottschalk2010} and \cite{Lupton2018}. However, when working with the model of \cite{Lupton2018} in practice, we found that many divergent samples were produced with the NUTS algorithm, suggesting the posterior samples have not converged. \cite{Dong2022} also encountered divergent samples in their study when using HMC, and opted to use Sequential Monte Carlo (SMC) instead to attempt to circumvent this problem. However, it is unclear that the samples from SMC are better converged than the samples from HMC, since SMC does not have access to the same divergent sample diagnostics as HMC. Furthermore, SMC appears to significantly increase computational time.

\cite{Cencic2015} developed a linear Bayesian data conciliation method that can be applied to MFA systems, and a subsequent paper \cite{Cencic2018} extended this method to include nonlinear constraints. These methods instead parametrise the model directly in terms of the flow variables, by partitioning the set of flow variables in the system into `free variables' $\bm{v_f}$ and `dependent variables' $\bm{v_d}$, where the mass balance equations can be expressed as a relationship between the free variables and the dependent variables. For example, in the linear case, $\bm{v_d}=-D\bm{v_f}-\bm{d}$ can be obtained via Gaussian elimination for some constant matrix $D$ and vector $\bm{d}$.
In both papers the methodology was tested on low dimensional, simulated examples, with Metropolis Hastings (MH) used as the sampling algorithm on the free variables, with the prior distribution $f(\bm{v_f)})$ of the free variables as the proposal distribution. However, even in low dimensional examples it was reported in \cite{Cencic2015} that the MH sampler can have a very low acceptance probability. Therefore it is unclear whether the method works well in high dimensional, underdetermined systems with hundreds of flow and change in stock variables, which is important since these systems are typical in MFA. In high dimensions, it becomes increasingly unlikely for proposals from the MH sampling algorithm to satisfy both the mass balance conditions and non negativity of flow variables. To see this, $\bm{v_d}=-D\bm{v_f}-\bm{d}$ does not guarantee every component of $\bm{v_d}$ will be positive for arbitrary fixed $D$ and $\bm{d}$. As the dimension of $\bm{v_d}$ increases, it becomes increasingly likely that randomly sampled proposal $\bm{v_f}$ during MH will lead to at least one component of $\bm{v_d}$ to be negative, causing the proposal to be in a region of zero posterior probability and the MCMC algorithm to be stuck at the current value. More generally, sampling from constrained posteriors is known to be challenging (\cite{Lan2023}), and HMC also have difficulties when encountering regions of zero posterior probability formed by the constraints (\cite{Hoffman2014}). 

\subsection{Scope of paper} \label{subsec: paperscope}

This paper continues the development of Bayesian methodology for material flow analysis. To address the aforementioned computational issues, we relax the mass balance conditions via a noise term. This reduces regions of zero posterior probability in the parameter space, making the posterior easier to sample using MCMC algorithms. We show computationally with an aluminium cycle case study that this leads the NUTS sampling algorithm to converge well in high dimensions. Simultaneously, the noise term can be interpreted as a way of modelling epistemic uncertainty in the system, which is likely to be present as MFA systems are simplified rather than perfect representations of reality (\cite{Schwab2016, Schwab2016other}). It is similar to the concept of `phantom flows', which is used to account for unexplainable mass imbalances in MFA studies (e.g. \cite{Reck2008} ). The variance of the noise term can be chosen to be small so good approximate mass balance is still achieved when there is high confidence in the system definition.

We introduce a child and parent process parametrisation framework to model systems with multiple layers of disaggregation in processes and flows, which is a common feature in material flow datasets (\cite{Myers2019} and \cite{Myers2019UMIS}) but has not been considered in previous Bayesian MFA studies. To this end we assign priors directly on flow mass and change in stock variables in the material system, while retaining the ability to incorporate ratio data between arbitrary flows. We demonstrate our method on a high dimensional aluminium material flow system where change in stock and disaggregation of processes and flows are simultaneously present. 

We illustrate how Bayesian posterior predictive checks can reveal inconsistencies in the data and aid hyperparameter selection. We also show the posterior distribution can inform data collection strategies by identifying which flow or changes in stock variables in the system retain the most uncertainty. 

Previous work have not investigated under what conditions estimates and uncertainty quantification produced by Bayesian MFA are reliable. This is important as strong theoretical guarantees which hold in low-dimensional parametric models, such as the Bernstein-von Mises theorem (\cite{Vaart1998}), can fail to hold in high-dimensional settings (\cite{Johnstone_2010}) such as the present MFA setting. We address this gap by examining how Bayesian MFA performs on a high dimensional aluminium case study under relatively weak assumptions, namely a weakly informative prior and significant data gaps. We also conduct a simulation study on a zinc cycle to examine the estimation accuracy and uncertainty quantification of the posterior distributions of our model from a frequentist perspective, which can be found in the supporting information.

\section{Methods} \label{sec: methods}

In this section, we present the details of our proposed Bayesian MFA methodology. Our model produces estimates and uncertainty quantification of all flow and change in stock variables of interest in any given material flow system, in the form of a posterior distribution, which mathematically combines prior domain knowledge and expert opinion with available data while simultaneously propagating uncertainty. 

A MFA analysis begins with the definition of a \textit{system diagram}, a graph like structure of nodes representing $processes$, where the material of interest can be stored as stock, and edges representing $flows$ of the material of interest between processes. Notably however, flows in both directions between any two processes are permitted. The system diagram will also contain a system boundary, which is used to describe the flows between the system and some external environment. This framework is quite broad, and allows processes within a system to not only represent a physical manufacturing process (such as components of a blast furnace), but also examples such as the Earth's lithosphere, the environment, or various usage outlets like households where the material of interest can be stored or flow in and out of. Similarly, the system's scope can range from a small supply chain, to a global flow of a metal such as zinc. The level of detail of the system diagram is chosen by the modeller to fit the scope of the problem being examined.

Often in MFA problems, certain processes in the system can be disaggregated into constituent subprocesses (see e.g. \cite{Myers2019UMIS} ). We define a $parent$ $process$ to be a process which contains subprocesses, and a $child$ $process$ be a process which contains no subprocesses instead. By definition, parent and child processes form a partition of the set of all processes in the system. The MFA practitioner should decide the level of disaggregation of each parent process according to their requirements, but parent processes where data is only available on some of its child processes may still be worth disaggregating to incorporate additional data into the model. 

The parent and child process structure is useful for modelling the disaggregation of processes. To see this, we assume the stocks and flows of any parent process can be expressed as a linear combination of its constituent child processes. Under this parent and child process framework, multiple levels of disaggregation of processes can be reduced to just two levels (the set of parent processes and the set of child processes), which greatly simplifies modelling of the MFA system. A simple example illustrating the parent and child process framework can be found in the supplementary information. 

\subsection{Formulation of the physical model} \label{subsec: physicalmodel}

Suppose there are $m$ child processes in the system, indexed by $P_0, P_1, \dots P_{m-1}$. Let $s_i(t)$ to be the stock variable associated with the process $P_i$, denoting the amount of stock in process $P_i$ at time $t$. In practice material flow data are typically recorded as the total amount of flow over a period time (such as on a monthly or yearly basis), so let $U_{j,i}$ represents the total amount of flow of the material of interest from process $j$ to process $i$ during the time period $t-\Delta t$ to $t$. Here $\Delta t$ represents the period during which the total amount of flow was reported, which can for example be in months or years. Note typically in MFA systems not all processes necessarily possess a stock variable, for example if the physical process which it is modelling does not contain a physical stock of the material of interest. Similarly, most processes will not receive flows from or flow to every other process, so the notation introduced in this paragraph such as $s_i$ and $U_{j,k}$ are understood to be over existing stocks and flow variables only. For each process $P_i$, we assume mass is conserved between its stock, inflows and outflows over the time period $t-\Delta t$ to $t$, which we formulate as:

\begin{equation} 
S_i=s_i(t)-s_i(t-\Delta t) = \sum_{j}^{}U_{j,i}-\sum_{k}^{}U_{i,k} 
\label{eq: massconservationintegrated}
\end{equation} 

The left hand side $S_i=s_i(t)-s_i(t-\Delta t)$ is simply the change in stock during the time period $t$ to $t-\Delta t$. In this paper we only consider stationary models at a snapshot of time $t$, so the variables of interest in the model are the $S_i$ and $U_{j,k}$. For more compact notation, let $\bm{S}$ be the vector of $q$ change in stock variables $S_i$ of the child processes in the system, and $\bm{U}$ a vector of $p-q$ flow variables $U_{j,k}$ of the child processes in the system (for a total of $p$ variables). Note because flows and stocks changes of parent processes can be expressed linearly in terms of its constituent child processes, conservation of mass for all child processes automatically implies conservation of mass for all parent processes. Similarly, Bayesian modelling only needs to be conducted on the child processes, as the posterior samples on flow and stocks change variables of parent processes can be obtained by simply summing the posterior samples of the constituent child processes. 

\subsection{Data structure} \label{subsec: Datastructure}

Data in MFA can in principle be any arbitrary function $f(\bm{S},\bm{U})$ of the variables of interest $\bm{S}$ and $\bm{U}$. 
However, data and mass balance conditions typically come in the following four forms, which we can express in terms of $\bm{S}$ and $\bm{U}$:

\medskip

1. Observations of changes in stock of child processes or observations of flows between child processes. For example, in \Cref{fig: aluminiumflowdiagram}, $S_0=-37.2$ Mt could be used to describe the change in stock of the `Reserves' process, while the flow from `Reserves' to `Mining' can be described by $U_{0,1}=37.2$ Mt. Here the process `Reserves' is labelled by $P_0$ and the process `Mining' by $P_1$. 

\medskip

2. Observations of changes in stock of parent processes, or observations of flows between processes where at least one process is a parent, which can be treated as the sum of multiple flows. For example, the combined flow of $9.8$ Mt in \Cref{fig: aluminiumflowdiagram} from `WasteManagement' to `Recycling' consists of `WasteManagement' as the origin process, and the destination processes are `Remelting' and `Refining', two of the subprocesses of `Recycling'. This flow can be represented by the sum of flows $U_{31,7}+U_{31,8}=9.8$ Mt. Here we used $P_{31}$ to denote `WasteManagement', and $P_{7}, P_{8}$ `Remelting' and `Refining' subprocesses of `Recycling' respectively. So for example, $U_{31,7}$ represents the flow from `WasteManagement' to `Remelting'. 

\medskip

3. Conservation of mass. For child process $i$ this is represented by equation \eqref{eq: massconservationintegrated}.

\medskip

4. Ratio data between flows or sums of flows, for instance transfer coefficients:

\begin{equation} \label{eq: ratiodata}
\frac{U_{i,j}}{{\sum_{k}^{}U_{i,k}}}=\alpha_{i,j}
\end{equation} 
where $\alpha_{i,j}$ is a known transfer coefficient of the flow from process $i$ to process $j$, and is defined as the ratio between the flow from process $i$ to process $j$, divided by the total outflow of process $i$. In case studies of this paper we only consider transfer coefficients of processes without a change in stock variable. In general, a change in stock can be split into two flows (flow into and out of stock), which allows transfer coefficients to be calculated. We also consider ratios between aggregated flows as an extension beyond standard transfer coefficients. 

Ratio data can also be alternatively parametrised linearly in the following way:

\begin{equation} \label{eq: ratiodatalinear} 
U_{i,j}-\alpha_{i,j}{\sum_{k}^{}U_{i,k}}=0
\end{equation} 

We note it is possible to conveniently formulate all the most common forms of data and mass balance conditions in MFA in terms of linear relationships between the change in stock variables $S_i$ and the flow variables $U_{j,k}$, even when the system contains disaggregation of processes. However, for a more general framework and to demonstrate our model can incorporate nonlinear data as well, we choose to parametrise ratio data as \Cref{eq: ratiodata} when evaluating the model on the aluminium case study in the Results section. 

Using more compact notation, the relationship between the flow and change in stock variables and the data and physical constraints can be represented by the following model:

\begin{align} 
\bm{Y}=\begin{bmatrix}
         X\bm{\theta} \\
         \bm{R}(\bm{\theta})
         \end{bmatrix} +\bm{\epsilon}, \quad
         \bm{\theta}=\begin{bmatrix}
           \bm{S} \\
           \bm{U} 
         \end{bmatrix}
\end{align} \label{eq: regressionequation}
where $\bm{\theta}$ is a concatenated vector of $\bm{S}$ and $\bm{U}$ of length $p$, representing all the child flow and change in stock variables the system. $\bm{Y}$ is a vector of length $n$ of observed values, or $0$ for mass balance conditions. $X$ is a design matrix representing linear data and mass balance conditions, and $\bm{R}(\bm{\theta})$ a vector representing nonlinear data. $\bm{\epsilon}$ is a random vector (of length $n$) representing uncertainty in the data, which could be caused by measurement or rounding errors. Note in the MFA setting, it is often the case that $n \ll p$.

The goal of any Bayesian model is to obtain posterior distributions over the variables of interest, in this case $\bm{\theta}$, in order to perform inference such as point estimation or uncertainty quantification. In the following section we describe how to construct the prior for our model, the form of the likelihood and how to obtain the posterior of $\bm{\theta}$ via Bayes's theorem.

\subsection{Bayesian model detail} \label{subsec: truncatedgaussianmodel}

Bayesian inference is a statistical framework in which Bayes's theorem, a fundamental result in probability theory, is used to update one's beliefs regarding parameters of interest $\bm{\theta}$ based on new data or evidence $\bm{Y}$. Mathematically the prior distribution $p(\bm{\theta})$ is used to express one's belief prior to seeing the data, and the likelihood function $p(\bm{Y}|\bm{\theta})$ used to express the probability of observing the data. The goal in Bayesian inference is to obtain the posterior distribution $p(\bm{\theta}|\bm{Y})$, which represents the state of one's updated beliefs after observing the data. This is done via Bayes's theorem:

\begin{equation} \label{eq: bayestheorem} 
p(\bm{\theta}|\bm{Y})=\frac{p(\bm{\theta})p(\bm{Y}|\bm{\theta})}{\int_{}^{} p(\bm{\theta})p(\bm{Y}|\bm{\theta}) d \bm{\theta} }
\end{equation}

Most posterior distributions do not have a closed form expression due not being possible to evaluate $\int_{}^{} p(\bm{\theta})p(\bm{Y}|\bm{\theta}) d \bm{\theta}$ in closed form. Markov Chain Monte Carlo (MCMC) methods are computational algorithms used to sample from distributions that do not have a closed-form, including posterior distributions. MCMC methods construct a Markov Chain that converges to the target distribution, by iteratively sampling from a proposal distribution and only accepting samples that satisfy a suitable criteria that suggests they could be feasibly generated from the target distribution.

\begin{figure}[h!]
\begin{subfigure}[b]{1.00\textwidth}
\includegraphics[width = 1.00\textwidth]{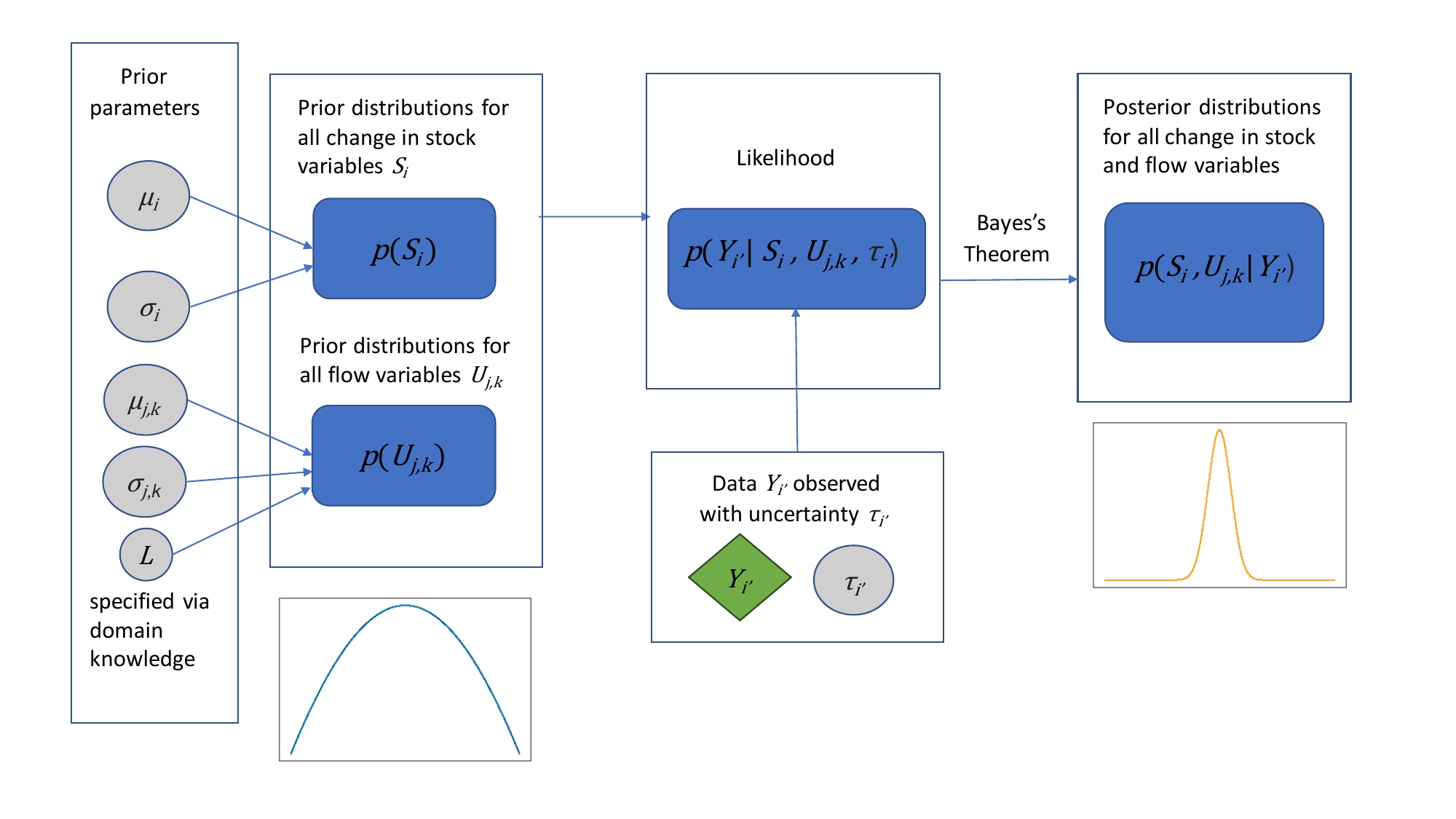}
\caption{}
\label{fig:modeldiagramsub}
\end{subfigure}
\begin{subfigure}[b]{1.00\textwidth}
\includegraphics[width = 1.00\textwidth, clip, trim = 0cm 5cm 0cm 3cm]{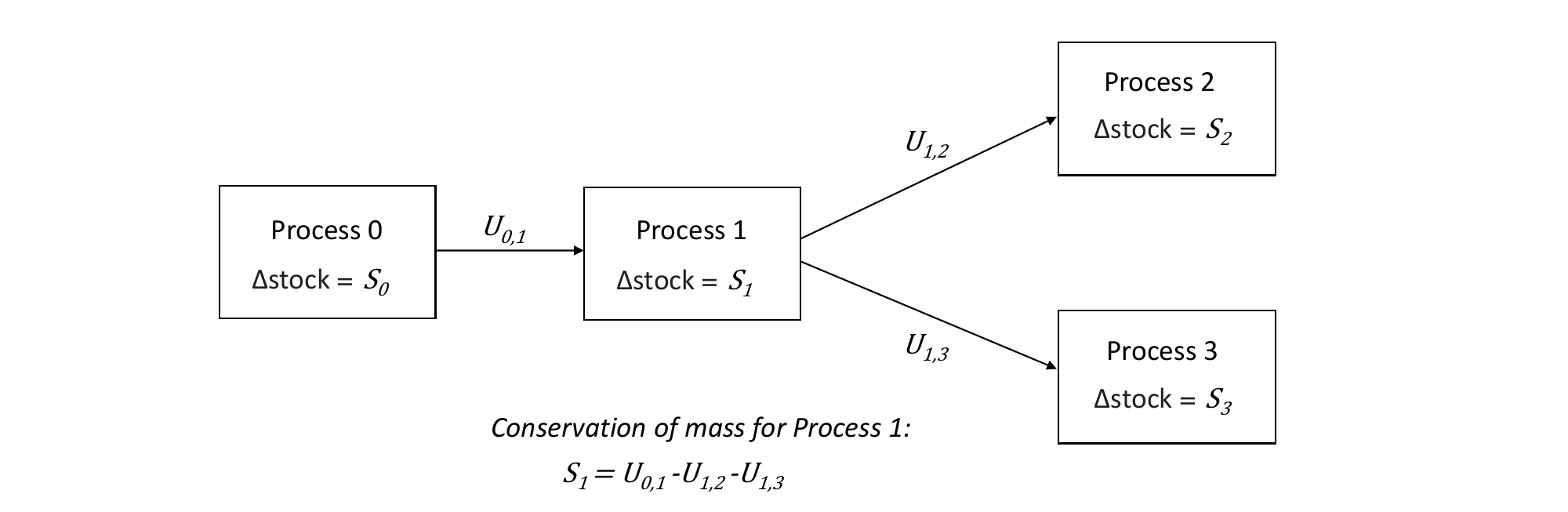}
\caption{}
\label{fig:modelsimpleexample}
\end{subfigure}
\caption{Schematic of the Bayesian model \textbf{(a)}, example of simple MFA system \textbf{(b)}. The circles coloured in grey represent noise parameters or prior hyperparameters, the diamond coloured in green represent the data and the rounded rectangles coloured in blue represent distributions over the variables of interest as well as the model likelihood. Definitions for each variable and parameter in this figure can be found in the Methods section. 
}
\label{fig: modeldiagram}
\end{figure}

For our Bayesian model, we consider normal priors for the change in stock variables $S_i$, which reflects the fact that the change in stock can be both positive or negative. For the flow variables $U_{j,k}$, we consider truncated normal priors for the flow variables on the positive interval $[0,L]$, where $L$ is chosen based on domain knowledge of the scale of flows inside the system being modelled, or simply chosen to be positive infinite if none is available. This is to ensure the posterior distribution of the flow variables have positive support, as flow quantities are always positive in reality. Furthermore, the normal and truncated normal distributions are flexible, in the sense that an informative prior distribution can be assigned by choosing the prior mode as a confident estimate (from expert knowledge), and choosing a small prior variance to create a narrow distribution around the prior mode. On the other hand, an uninformative prior distribution can be assigned by choosing a large prior variance around a rough guess for the prior mode instead.

\begin{align}
S_i \sim \mathcal{N}(\mu_i, \sigma_i^2),\quad  U_{j,k} \sim \mathcal{TN}(\mu_{j,k}, \sigma_{j,k}^2, 0, L)
\end{align} \label{eq: truncatednormalprior}
where $\mu_i$ and $\sigma_i^2$ are the prior mean and variance of $i$th change in stock variable $S_i$, respectively, and $\mu_{j,k}$ and  $\sigma_{j,k}^2$ the corresponding prior hyperparameters respectively for the truncated normal distribution for the flow variable $U_{j,k}$, representing the flow quantity from process $j$ to process $k$. We also assume prior variables are independent, so the overall prior distribution $p(\bm{\theta})$ over the change in stock and flow variables is of the form:

\begin{align}
p(\bm{\theta}) \propto \prod_{i}^{} \frac{1}{\sigma_i}\exp(\frac{-(S_i-\mu_i)^2}{2\sigma_i^2}) \prod_{(j,k)}^{} \frac{1}{\sigma_{j,k}}\exp(\frac{-(U_{j,k}-\mu_{j,k})^2}{2\sigma_{j,k}^2}) \mathrm{I}_{U_{j,k} \in [0,L]}
\end{align} \label{eq: priorform}
where $\mathrm{I}_{U_{j,k} \in [0,L]}$ is an indicator function, and the products are over the change in stock variables $S_i$ and flow variables $U_{j,k}$.

For the likelihood, we again use a truncated normal likelihood for any data on observed flow values to ensure the data generated by the likelihood is plausible (since flow data should be strictly positive), and a normal likelihood for data on observed change in stock values and general nonlinear data. We include an additional normally distributed noise term in the mass balance conditions mainly for practical reasons: the noise term reduces the region of zero posterior probability, which makes sampling from the posterior computationally much more tractable. Additionally, the noise term has the interpretation of modelling epistemic or systematic uncertainty in the MFA system. In theory, every process should be exactly mass conserved if the system definition and diagram are a perfect representation of the underlying real system being modelled. However, in reality the system diagram and definition are simplified and approximate models of reality, and so it is reasonable to account for uncertainty in the underlying system itself, which can be modelled through a noise term in the mass balance conditions. In practice, flows reported in MFA studies are rarely exactly mass balanced, so likewise a Bayesian MFA approach does not need to produce perfectly mass balanced flows to provide useful estimates.

\begin{align}
Y_{i'}|\bm{\theta} & \sim \mathcal{N}(\bm{x_{i'}}^\top\bm{\theta}, \tau_{i'}^2) \quad \text{for observations} \ Y_{i'} \ \text{on change in stock data} \\
Y_{j'}|\bm{\theta} & \sim \mathcal{TN}(\bm{x_{j'}}^\top\bm{\theta}, \tau_{j'}^2, 0, \infty) \quad \text{for observations} \ Y_{j'} \ \text{on flow data} \\
Y_{k'}|\bm{\theta} & \sim \mathcal{N}(R(\bm{\theta})_{k'}, \tau_{k'}^2) \quad \text{for observations} \ Y_{k'} \ \text{on nonlinear data} \\
Y_{l'}|\bm{\theta} & \sim \mathcal{N}(\bm{x_{l'}}^\top\bm{\theta}, \tau_{l'}^2) \quad \text{for} \ Y_{l'}=0 \ \text{on mass balance conditions} 
\end{align} \label{eq: truncatednormallikelihood}

Here $Y_{i'}$ is the $i'$th row or entry of the observation vector $\bm{Y}$, and $\bm{x_{i'}}^\top$ the $i'$th row of the design matrix $X$, $R(\bm{\theta})_{k'}$ the $k'$th row of the vector $\bm{R}(\bm{\theta})$ and $\tau_{i'}^2$ representing variance of the noise of the $i'$th observation, which are assumed to be independent. The overall likelihood $p(\bm{Y}|\bm{\theta})$ is therefore of the form:

\begin{align}
p(\bm{Y}|\bm{\theta})=\prod_{i'}^{} p(Y_{i'}|\bm{\theta})\prod_{j'}^{} p(Y_{j'}|\bm{\theta})\prod_{k'}^{} p(Y_{k'}|\bm{\theta})\prod_{l'}^{} p(Y_{l'}|\bm{\theta})
\end{align} \label{eq: likelihoodform2}
where
\begin{align}
p(Y_{i'}|\bm{\theta}) & \propto \frac{1}{\tau_{i'}}\exp(\frac{-(Y_{i'}-\bm{x_{i'}}^\top\bm{\theta})^2}{2\tau_{i'}^2})  \\
p(Y_{j'}|\bm{\theta}) & \propto \frac{1}{\tau_{j'}}\exp(\frac{-(Y_{j'}-\bm{x_{j'}}^\top\bm{\theta})^2}{2\tau_{j'}^2}) \mathrm{I}_{Y_{j'} \in [0,\infty)} \\
p(Y_{k'}|\bm{\theta}) & \propto \frac{1}{\tau_{k'}}\exp(\frac{-(Y_{k'}-R(\bm{\theta})_{k'})^2}{2\tau_{k'}^2}) \\
p(Y_{l'}|\bm{\theta}) & \propto \frac{1}{\tau_{l'}}\exp(\frac{-(Y_{l'}-\bm{x_{l'}}^\top\bm{\theta})^2}{2\tau_{l'}^2})  
\end{align} \label{eq: likelihoodformsplit}

Given the prior and likelihood, we can use Bayes's Theorem \ref{eq: bayestheorem} to obtain the posterior distribution $p(\bm{\theta}|\bm{Y})$. However, the posterior induced by this model does not admit an analytical form, and so we employ the No-U-Turn Sampler (NUTS) algorithm of \cite{Hoffman2014} to sample from the posterior, implemented via the \texttt{PyMC3} library \cite{Salvatier2016} in Python. NUTS is a Hamiltonian Monte Carlo that achieves increased sampling efficiency over traditional MCMC methods, such as Metropolis Hastings, by exploiting the gradient information of the target distribution to generate more informed sample proposals and explore the target distribution more efficiently. This is especially important in high dimensions (which applies to many MFA systems), since the probability mass of the target distribution are more likely concentrated in smaller regions, which is inefficient to explore via Metropolis Hastings due to random walk behaviour. 

\subsection{Posterior predictive checks} \label{sec: posteriorpredictivechecks}

For Bayesian modelling of MFA systems, we recommend performing posterior predictive checks to verify whether data generated by the model is similar to the observed data and adequately mass conserved, which gives some assurance that the prior, model and parameters chosen are sensible. Here we briefly describe the method of posterior predictive checking outlined in Chapter 6 of \cite{Gelman2013}. Recall that in our model, the observation vector $\bm{Y}$ represents the observed data (such as on flows or change in stocks), as well as physical conditions such as mass balance, and $\bm{\theta}$ the vector of parameters. Let $\bm{Y^{rep}}$ be replicated data that could have been observed, the distribution of $\bm{Y^{rep}}$ given the observed data $\bm{Y}$, also known as the posterior predictive distribution, is given by:

\begin{equation}
p(\bm{Y^{rep}} | \bm{Y})=\int_{}^{} p(\bm{Y^{rep}} | \bm{\theta})p(\bm{\theta} | \bm{Y}) \,d \bm{\theta}
\label{eq: posteriorpredictiveintegral}
\end{equation} 
 
Typically, the check is done on suitable scalar test quantities $T(\bm{Y}, \bm{\theta})$, chosen based on the real problem being modelled. The test quantity of the replicated data $T(\bm{Y^{rep}}, \bm{\theta})$ is compared with the test quantity of the observed data $T(\bm{Y}, \bm{\theta})$ through statistical tests or graphical checks to look for systematic discrepancies between the simulated and originally observed data. For our Bayesian material flow analysis model, we choose the test quantities to be each individual observed data and mass balance conditions of child processes in the system, in other words we choose test quantities $T_i(\bm{Y}, \bm{\theta})=Y_i$ for each $i$. This ensures we minimally check using the posterior predictive distribution that the model is consistent with the existing data as well as mass balance of child processes. One way to compare the observed data with the posterior predictive distribution is to calculate the Bayesian posterior predictive p-values for the test statistic, in our case the marginal observations $Y_i$, which are given by 

\begin{equation}
pval_i=P(Y_i^{rep} \geq Y_i | \bm{Y})=\int_{}^{}\int_{}^{} \mathrm{I}_{Y_i^{rep} \geq Y_i} p(\bm{Y^{rep}} | \bm{\theta})p(\bm{\theta} | \bm{Y}) \,d\bm{Y^{rep}} d \bm{\theta}
\label{eq: posteriorpredictivepvalue}
\end{equation} 

Very large or small p-value (e.g. greater than 0.95 or smaller than 0.05 as suggested by \cite{Gelman2013}) suggests the observed test quantity is unlikely to be replicated in repeated experiments if the model were true, implying there is inconsistency between the model and the data. We also calculate the posterior predictive marginal $95\%$ highest density interval for each $Y_i | \bm{Y}$ to see if they contain the observed values $Y_i$ (which include the conservation of mass conditions where $Y_i=0$).

In practice, the integrals in \Cref{eq: posteriorpredictiveintegral} and \Cref{eq: posteriorpredictivepvalue} are analytically intractable, so we again approximate the posterior predictive distribution via sampling. Specifically we simulate one sample of $\bm{Y^{rep}}$ from the posterior predictive distribution for each posterior sample of $\bm{\theta}$, and we approximate p-values of \Cref{eq: posteriorpredictivepvalue} by checking the proportion of the posterior predictive samples of $Y_i^{rep}$ that exceed the observed value $Y_i$, for each $i$.

\subsection{Hyperparameter and noise parameter selection} \label{sec: parameterselection}

In general, the prior hyperparameters $\mu_i$, $\mu_{j,k}$ should be specified through domain knowledge to reflect the modeller's best estimate of the stock change and flow variables apriori, and $\sigma_i^2$, $\sigma_{j,k}^2$ chosen to reflect prior uncertainty, the less confident the estimates for $\mu_i$, $\mu_{j,k}$, the larger $\sigma_i^2$, $\sigma_{j,k}^2$. 

The noise variance parameters $\tau_{i'}^2$ should also ideally be chosen to reflect uncertainty in the data, and in the case the mass balance conditions, epistemic uncertainty in the system definition. We recommend using posterior predictive checking to help select suitable noise parameter values, especially if no knowledge of the degree of data uncertainty is available. One can start with a small choice of standard deviation parameters (such as $10\%$ of the observed data value, and a small constant for the mass balance conditions), run the model and conduct posterior predictive checks, and identify the data points and mass balance conditions which exhibit extreme Bayesian p-values. The standard deviation parameter for those data points or mass balance conditions should be increased and the model rerun until no extreme Bayesian p-values remain. 

Full details of hyper parameters choice in the case studies examined in the Results section can be found in the supporting information. We give examples of how to specify prior hyperparameters in the case where there is weakly informative domain knowledge where some flows are known to the nearest order of magnitude, as well as flows where there is no prior knowledge available. 

\section{Results} \label{sec: results}

In this section we demonstrate our Bayesian MFA method on an aluminium cycle, containing significant disaggregations of processes and flows. We evaluate our model on the aluminium cycle under a weakly informative prior for two different levels of data availability and compare the results. We also present posterior predictive checks to identify inconsistencies between the model and data, and processes which are not mass balanced by the available data.

\subsection{Aluminium cycle} \label{sec: aluminiumcasestudy}

\begin{figure}[h!]
\centering
\includegraphics[width = \textwidth]{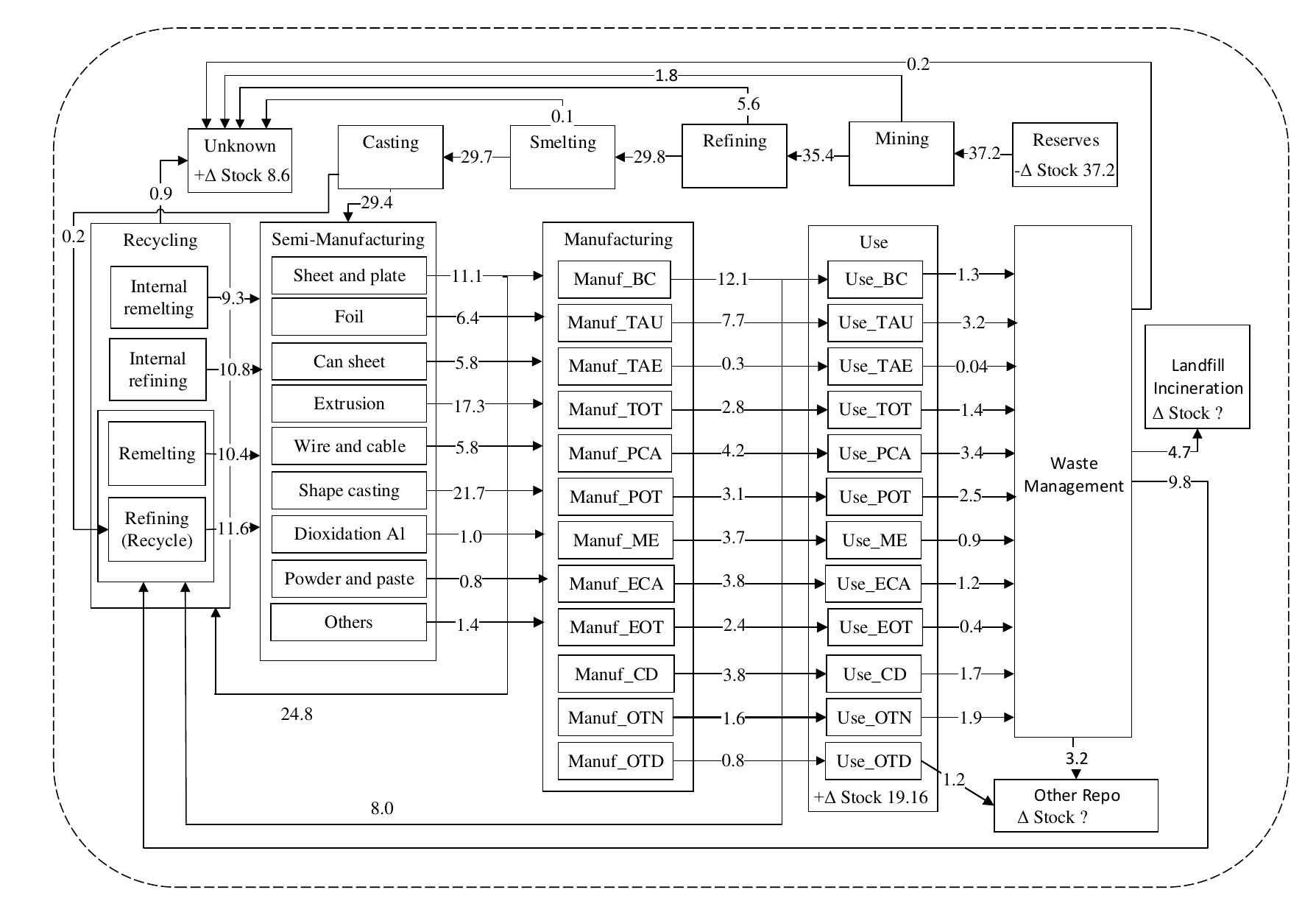}
\caption{Global anthropogenic metallurgical aluminium cycle in 2009, adapted from Figure 1 of \cite{Liu2013}. Mass of aluminium is measured in megatonnes (Mt). 
}
\label{fig: aluminiumflowdiagram}
\end{figure}

We evaluate our model on the global anthropogenic metallurgical aluminium cycle in 2009 from \cite{Liu2013}. The associated system diagram \Cref{fig: aluminiumflowdiagram} is adapted from Figure 1 of \cite{Liu2013}. The aluminium cycle contains significant disaggregations of processes. For example, the `Use' Process contains subprocesses such as `Use\_BC' (building and construction) and `Use\_ME' (machinery and equipment) representing different use product categories of aluminium. Furthermore, for aggregated processes such as `Manufacturing' and `Recycling', \Cref{fig: aluminiumflowdiagram} only displays data on aggregated flows and change in stocks. For example, the flow $12.1$ Mt from `Manufacturing' is a combined flow to `Use\_BC' and the subprocesses `Remelting' and `Refining' of `Recycling'. So it is not clear from \Cref{fig: aluminiumflowdiagram} alone in what proportion this flow should split among the constituent subprocesses. In tables S5, S6 and S8 of the supplementary information, \cite{Liu2013} provides ratios specifying how certain aggregated flows should split into constituent subflows (known as transfer coefficients), specifically for the aggregate flows from `Semi-manufacturing' to `Manufacturing' and `Recycling', and from `Manufacturing' to `Use' and `Recycling'. For the purposes of testing our model, we treat the values presented in Figure 1 of \cite{Liu2013} and transfer coefficients in tables S5,S6,S6 of the supplementary information as data. We also henceforth use `reported value' to refer to any flow and change in stock values in Figure 1 of \cite{Liu2013}, as well as values of disaggregated flow and change in stock values calculated through the transfer coefficients in the supplementary material of \cite{Liu2013}.

We evaluate the model under two different scenarios. In scenario A, we deliberately withhold the transfer coefficients in the supplementary information of \cite{Liu2013} from the model, and only use the data displayed in \Cref{fig: aluminiumflowdiagram}. Instead, where transfer coefficients are available to calculate the value of the disaggregated flow, we set the prior mode of the disaggregated flow variable to the nearest power of $10$ of the reported value, otherwise an uninformative prior with mode $1.0$ Mt is used. The purpose of scenario A is to see if our Bayesian MFA model can still produce useful estimates of flows and changes in stock under a weakly informative prior with significant amount of missing data. Scenario A also mimics a situation that could be applicable to many MFA problems, where data is available on a aggregated/parent level but not the disaggregated/child level, and a rough estimate like the order of magnitude of the flows and changes in stock on the disaggregated level is obtained from surveying domain experts or approximate calculations, which can be used to construct a weakly informative prior. In scenario B, the same weakly informative prior is used, but the flow ratios in the supplementary information of \cite{Liu2013} are provided to the model as well. Scenario B mimics a situation towards the end of an MFA analysis, when data are available for most of the flow and change in stock variables in the system. In both scenarios, we assume a low degree of epistemic uncertainty in the system diagram and set the standard deviation of mass balance conditions to a constant $0.05$ Mt. This choice leads to well mass balanced posterior means and samples of the flow and change in stock variables, and we provide an analysis of posterior mass balance conditions in the supporting information (section S5) for full detail. Full detail of prior hyperparameters can also be found in the supporting information (section S4).

\begin{figure*}[h!]{
\centering
\includegraphics[width = 0.48\textwidth]{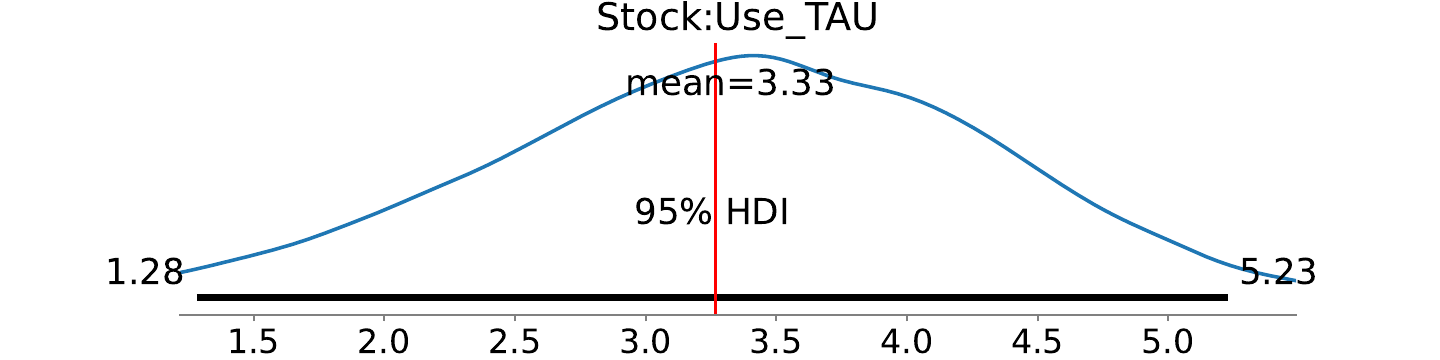}
\includegraphics[width = 0.48\textwidth]{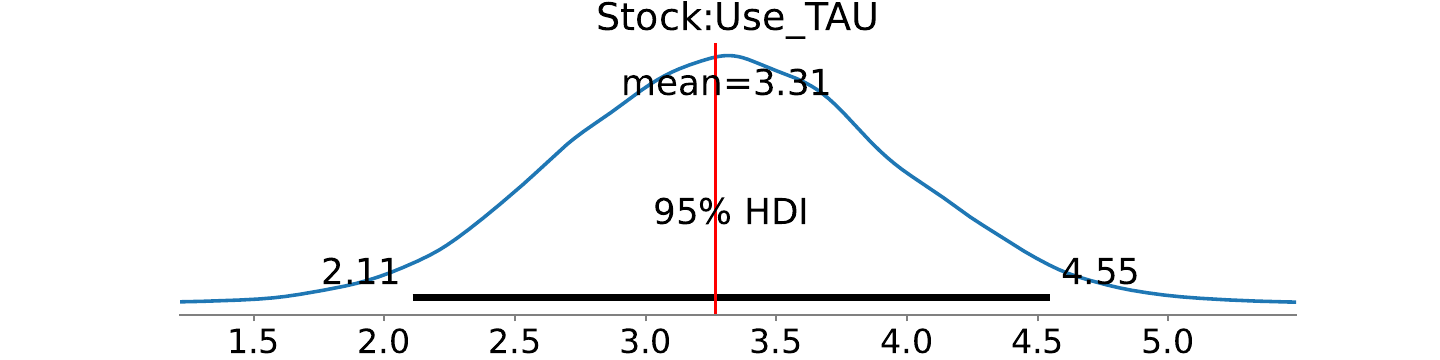}
\includegraphics[width = 0.48\textwidth]{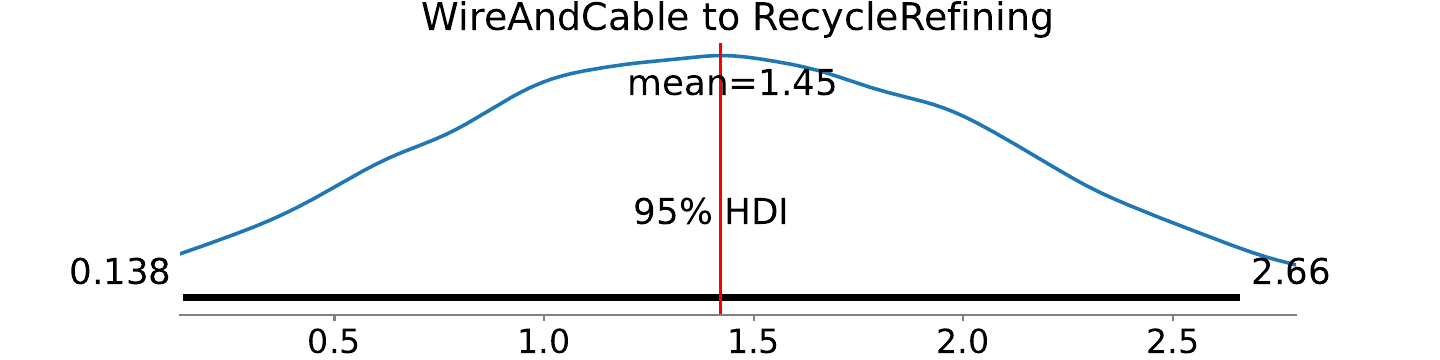}
\includegraphics[width = 0.48\textwidth]{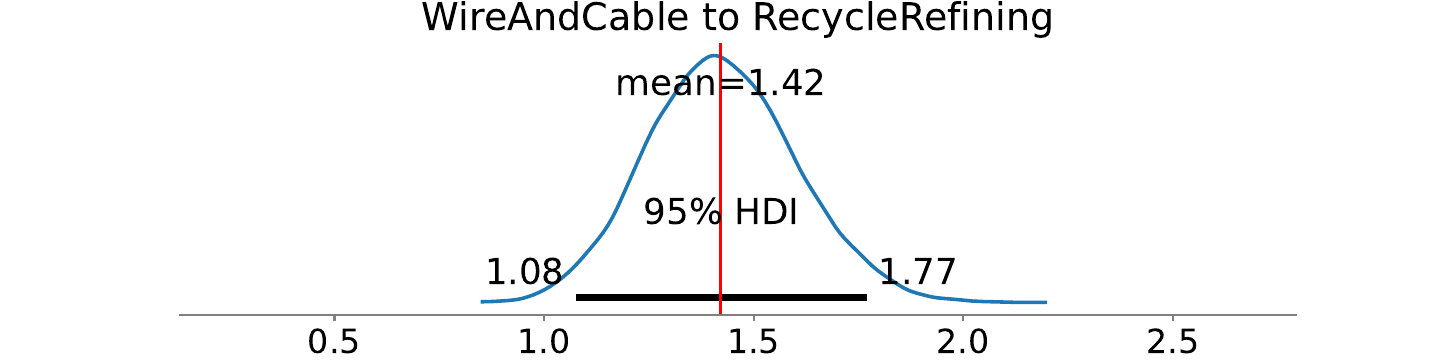}
\includegraphics[width = 0.48\textwidth]{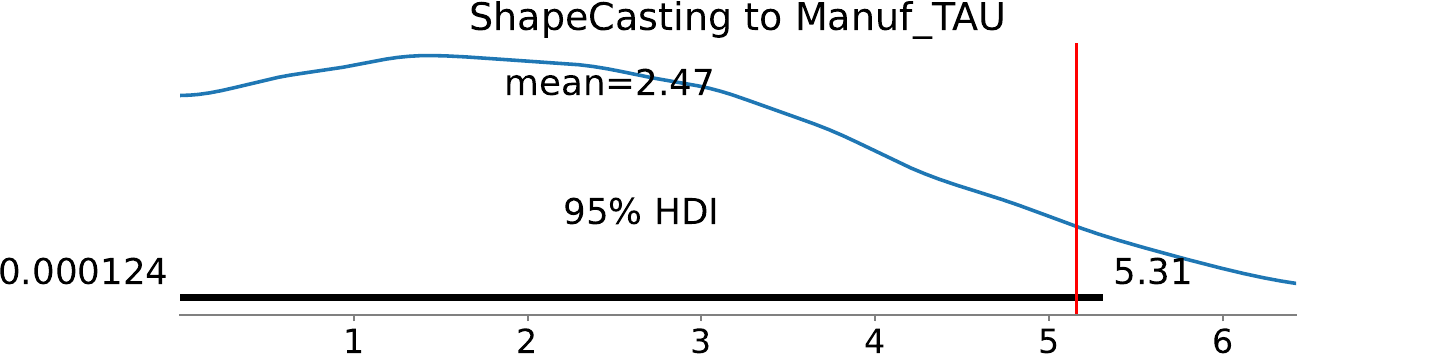}
\includegraphics[width = 0.48\textwidth]{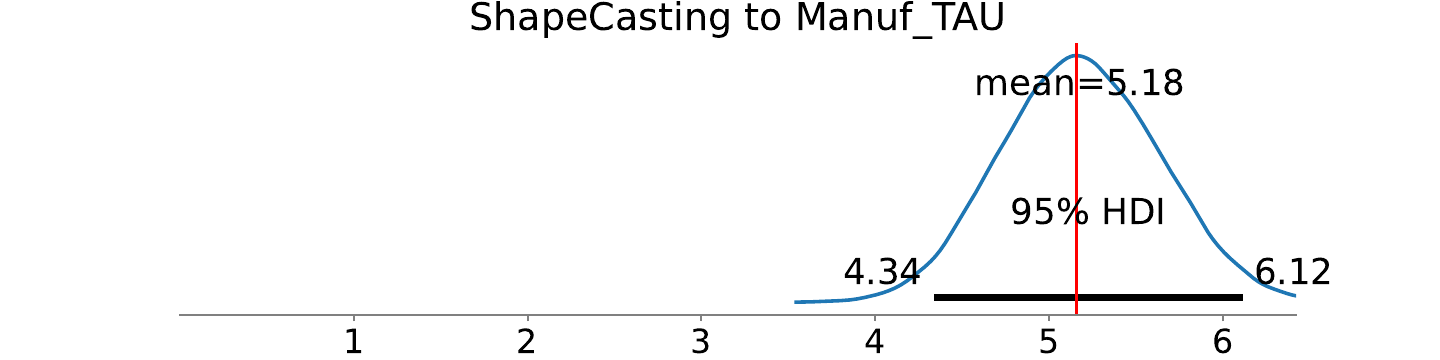}
\includegraphics[width = 0.48\textwidth]{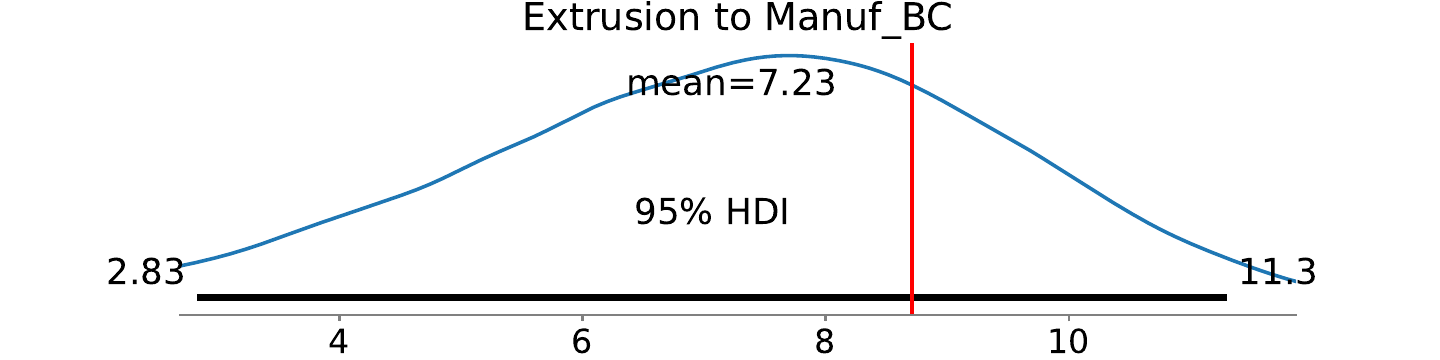}
\includegraphics[width = 0.48\textwidth]{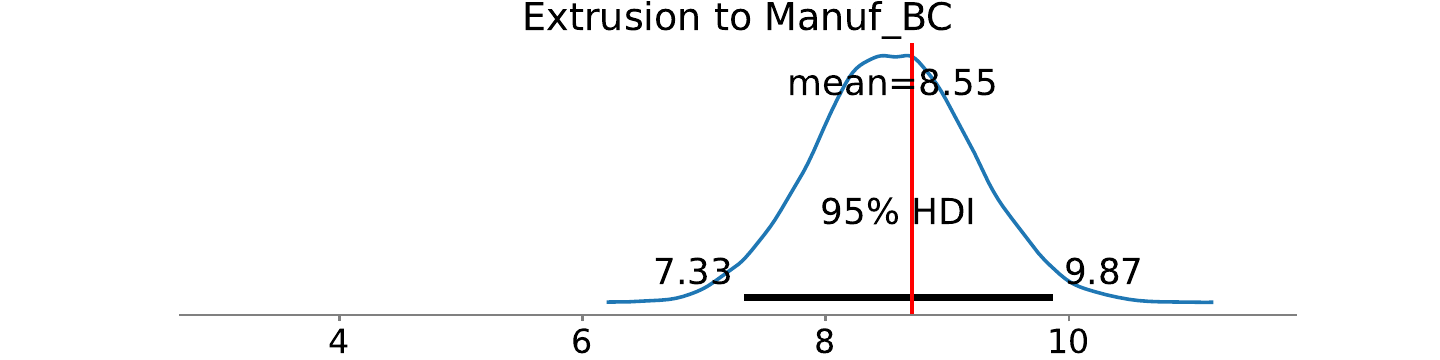}
\includegraphics[width = 0.48\textwidth]{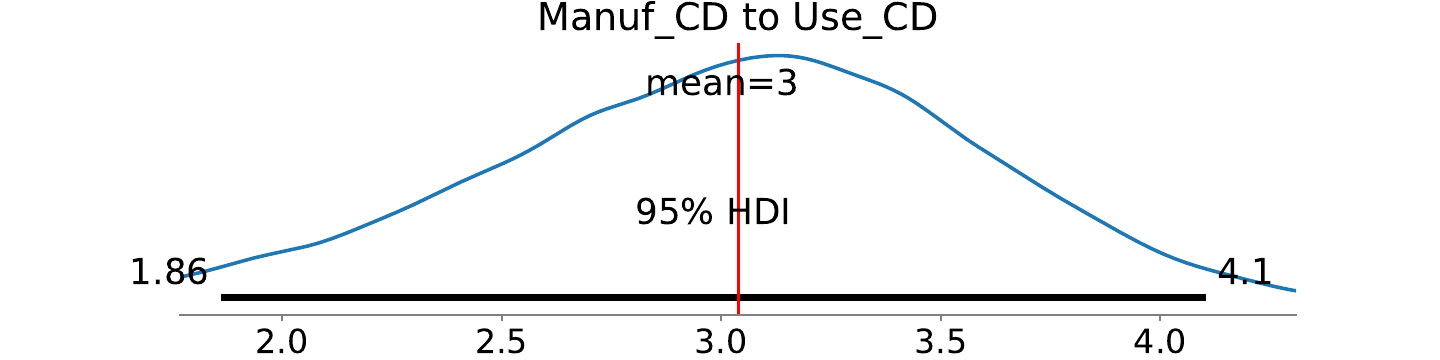}
\includegraphics[width = 0.48\textwidth]{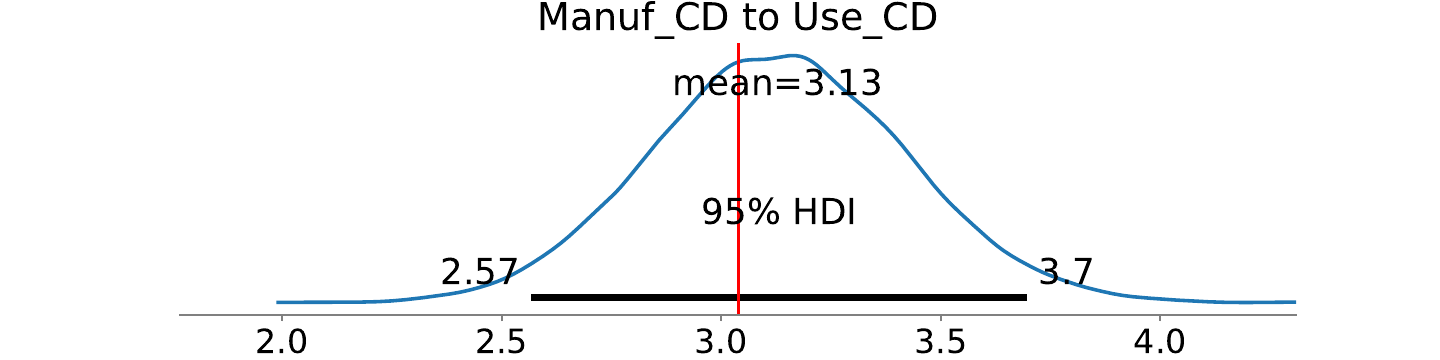}
\includegraphics[width = 0.48\textwidth]{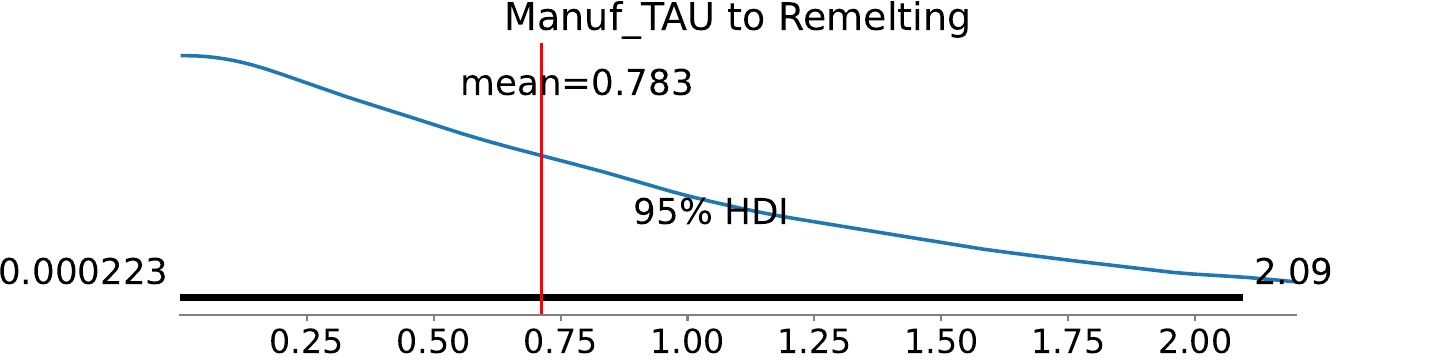}
\includegraphics[width = 0.48\textwidth,clip,trim = -1.7cm  0cm 0cm 0cm]{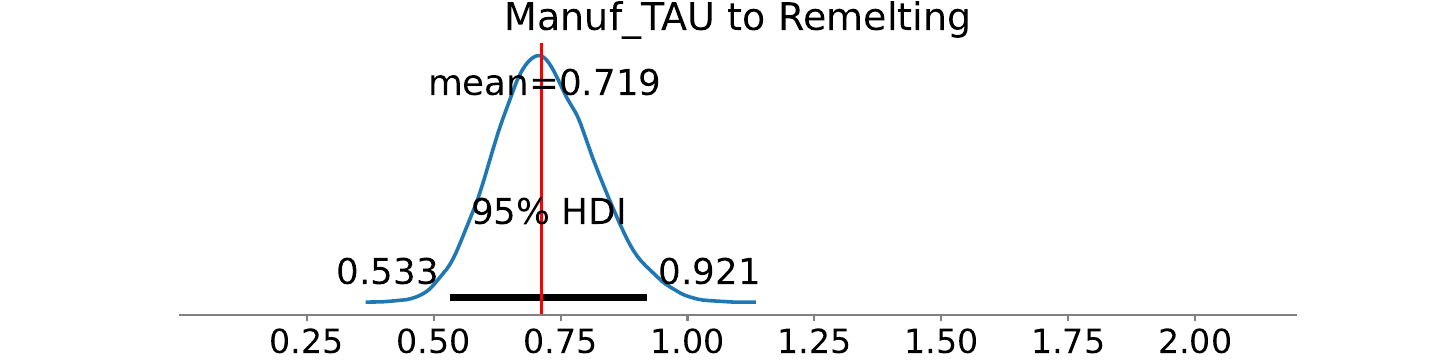}
}
\caption{Marginal posterior distributions for a selection of flow and change in stock variables of the aluminium dataset. For each flow or change in stock variables we display both the marginal posterior for scenario A (on the left) and for the scenario B (on the right). Each marginal posterior plot displays the mean and the 95 \% highest density interval (HDI). The red vertical line in each graph represents the reported value of the variable, calculated using transfer coefficients provided in the supplement of \cite{Liu2013}. Underlying data for this figure can be found at https://github.com/jwang727/BayesianMFA }
\label{fig: aluminiumposteriorplots}
\end{figure*}

\Cref{fig: aluminiumposteriorplots} displays a selection of marginal posterior distributions for flow and change in stock variables of disaggregated/child processes in the aluminium cycle. Only a selection is
displayed here for brevity as there are around $180$ flow or change in stock variables in the model. Under scenario A, the marginal posteriors are relatively more biased away from the reported value, which is not unexpected as data on the disaggregated level was withheld from the model in scenario A. Nevertheless, all of the reported values of flows and change in stock (when they are available in the supplement of \cite{Liu2013}) are contained within the $95\%$ posterior marginal highest density intervals (HDI). A small number of reported values are near the edge of HDI like the flow `ShapeCasting to Manuf\_TAU'. This could be caused by the reported value (around $5.1$ Mt) being close to the middle of the nearest orders of magnitudes ($1.0$ and $10.0$ Mt), which is more difficult for a prior based on the nearest order of magnitude ($1.0$ Mt in this case) to capture without actual data on the flow. Overall under scenario A, the posterior estimates and uncertainty quantification for disaggregated flows or change in stock variables obtained under still captures the reported values reasonably well despite not given data on any disaggregated flows. With the addition of transfer coefficient data under scenario B, the posterior marginal distributions generally possess narrower HDI compared to scenario A and centred much more around the reported value, and again the reported values are contained within the $95\%$ posterior HDI. Quantitatively, the average length of the $95\%$ posterior HDI over all flow and change in stock variables under scenario A is $2.15$ Mt, while under scenario B is $1.46$ Mt. 

In both scenarios it took the NUTS algorithm around 90 minutes to generate $24000$ posterior samples across two chains on an Intel i5-1145G7 CPU, 2.6GHz. The traceplots and convergence checks suggests the NUTS algorithm has converged and we did not observe any divergent samples. A selection of traceplots and further diagnostics can be found in section S2 of the supporting information. Posterior pairplots illustrating posterior correlation between a selection of flow variables can be found in section S3. In section S6 we include a simulation study on a zinc cycle to examine the estimation accuracy and uncertainty quantification of our model from a frequentist perspective. We demonstrate on a zinc cycle that incorporating even a weakly informative prior can significantly reduce estimation error, and the posterior credible intervals can consistently contain the true value of flows and stock changes and be interpreted as confidence intervals. 


\subsection{Posterior predictive checks on aluminium model} \label{sec: additionalaluminium}

\begin{figure*}[h!]{
\begin{subfigure}[b]{0.24\textwidth}
\includegraphics[width = 1.00\textwidth]{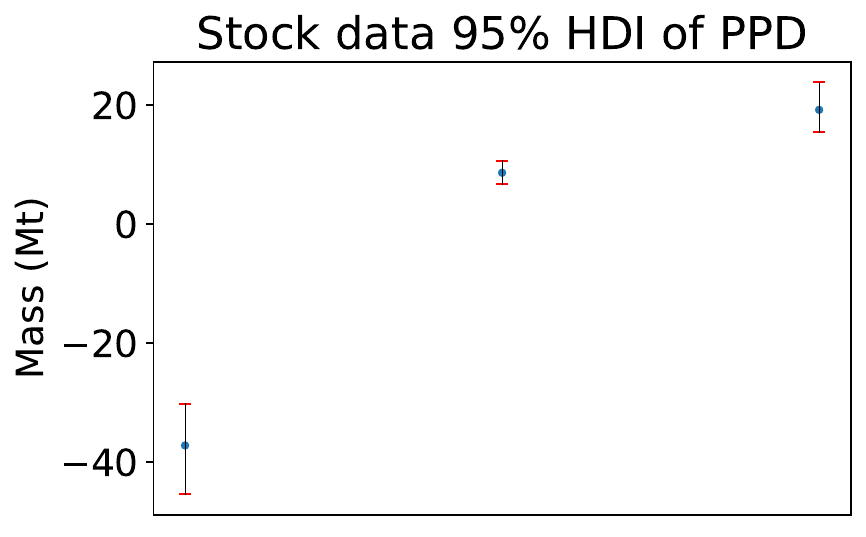}
\caption{}
\label{fig:stockpredictiviehdi}
\end{subfigure}
\begin{subfigure}[b]{0.24\textwidth}
\includegraphics[width = 1.00\textwidth]{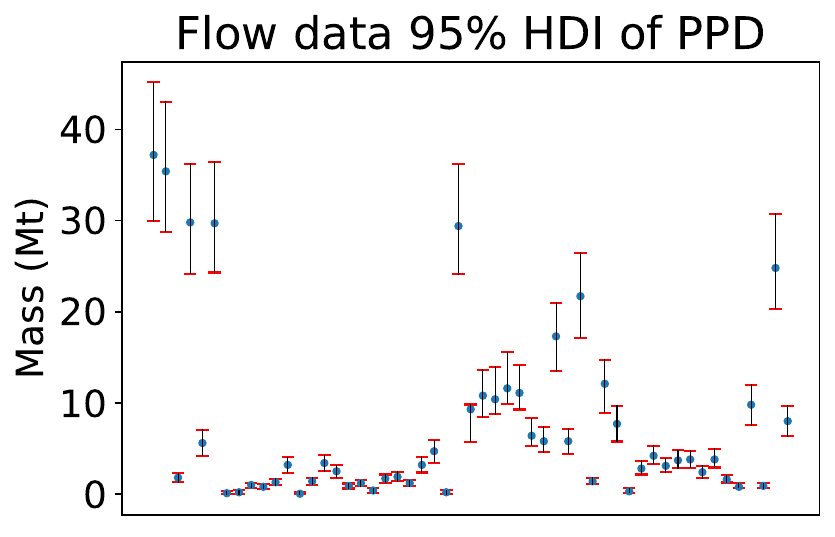}
\caption{}
\label{fig:flowpredictiviehdi}
\end{subfigure}
\begin{subfigure}[b]{0.24\textwidth}
\includegraphics[width = 1.00\textwidth]{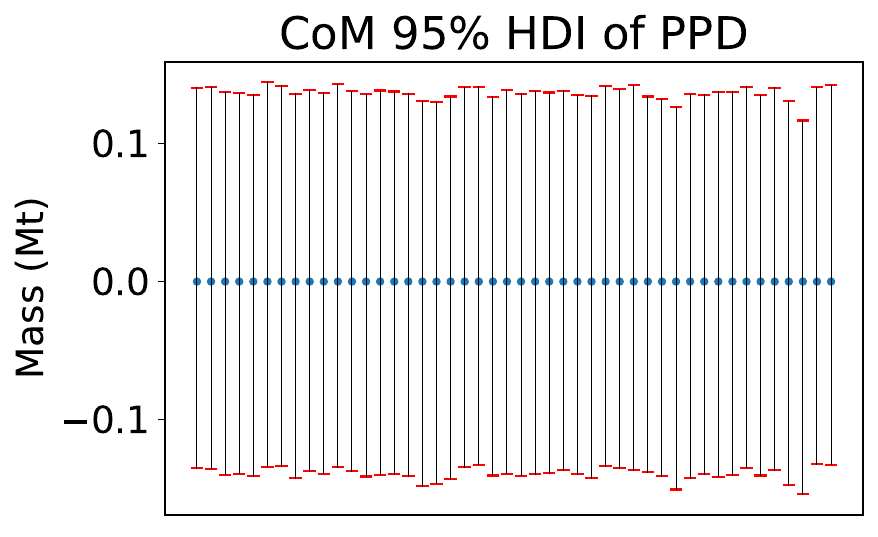}
\caption{}
\label{fig:CoMpredictiviehdi}
\end{subfigure}
\begin{subfigure}[b]{0.24\textwidth}
\includegraphics[width = 1.00\textwidth]{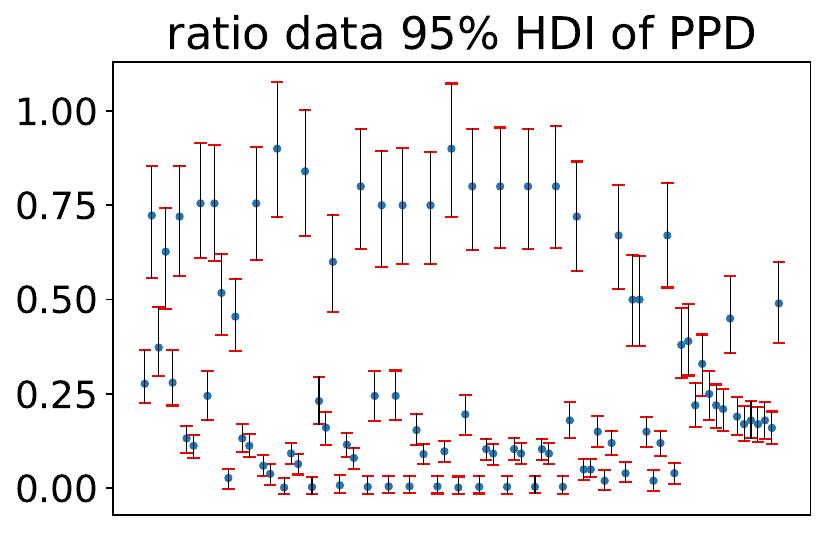}
\caption{}
\label{fig:ratiopredictiviehdi}
\end{subfigure}
\begin{subfigure}[b]{0.24\textwidth}
\includegraphics[width = 1.00\textwidth]{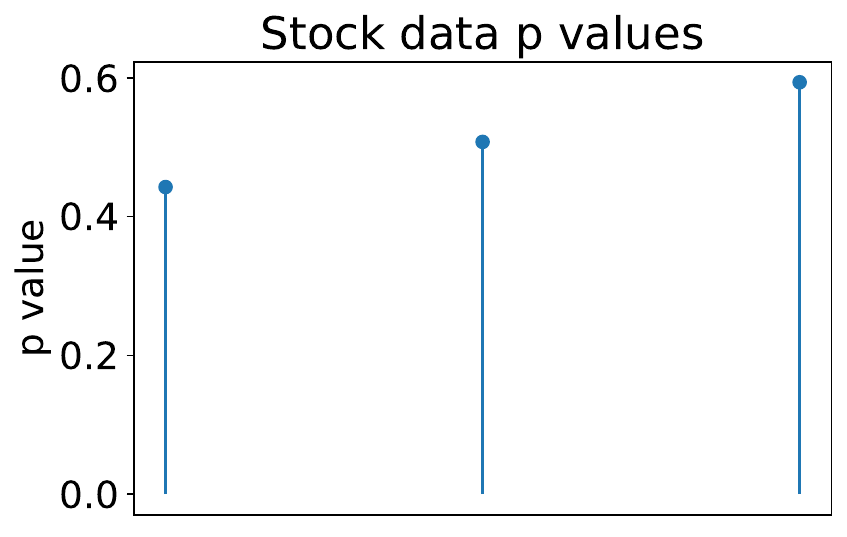}
\caption{}
\label{fig:stockpvalues}
\end{subfigure}
\begin{subfigure}[b]{0.24\textwidth}
\includegraphics[width = 1.00\textwidth]{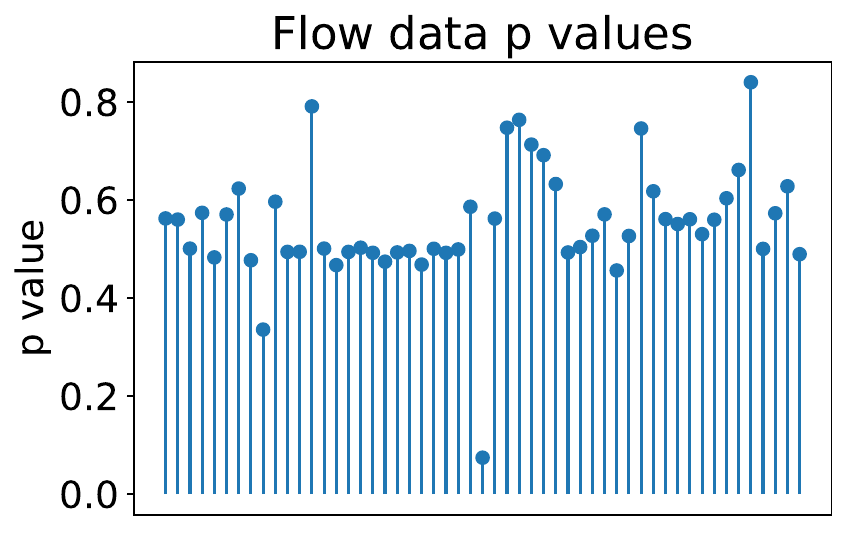}
\caption{}
\label{fig:flowpvalues}
\end{subfigure}
\begin{subfigure}[b]{0.24\textwidth}
\includegraphics[width = 1.00\textwidth]{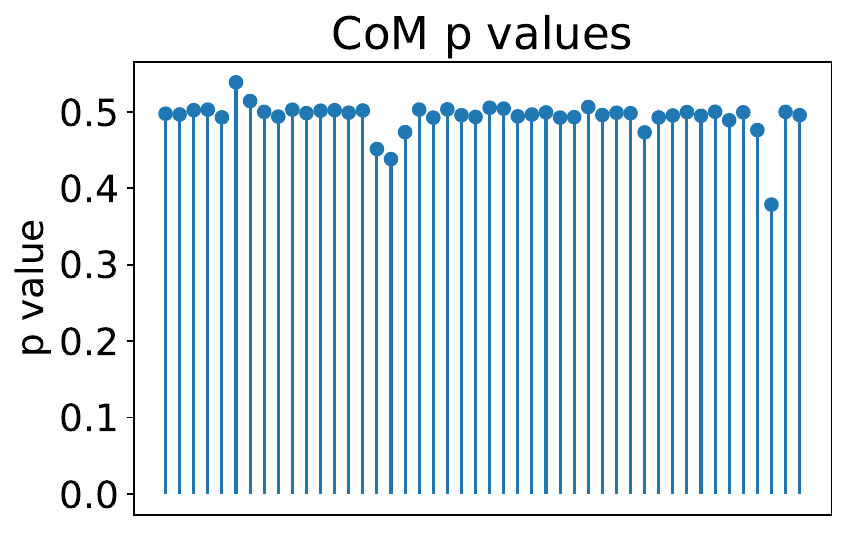}
\caption{}
\label{fig:CoMpvalues}
\end{subfigure}
\begin{subfigure}[b]{0.24\textwidth}
\includegraphics[width = 1.00\textwidth]{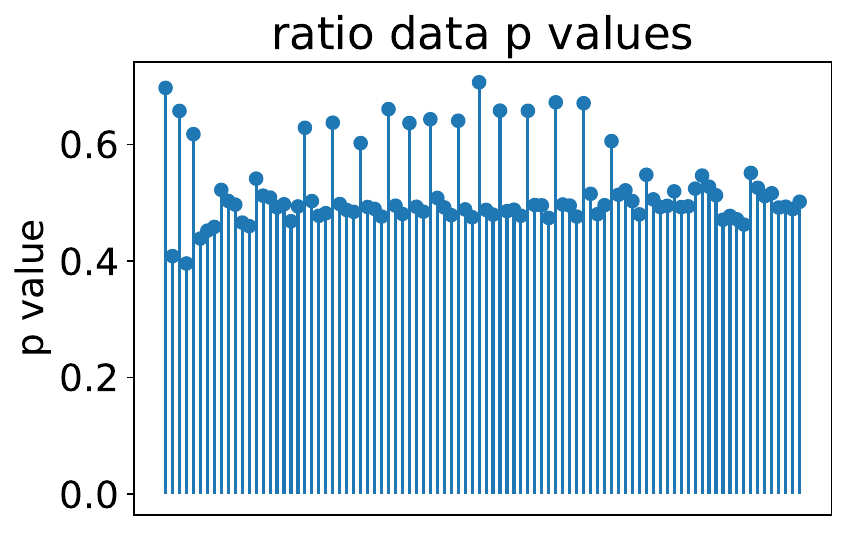}
\caption{}
\label{fig:ratiopvalues}
\end{subfigure}
\begin{subfigure}[b]{0.48\textwidth}
\includegraphics[width = 1.00\textwidth,clip,trim = 0cm  0cm 0cm 0cm]{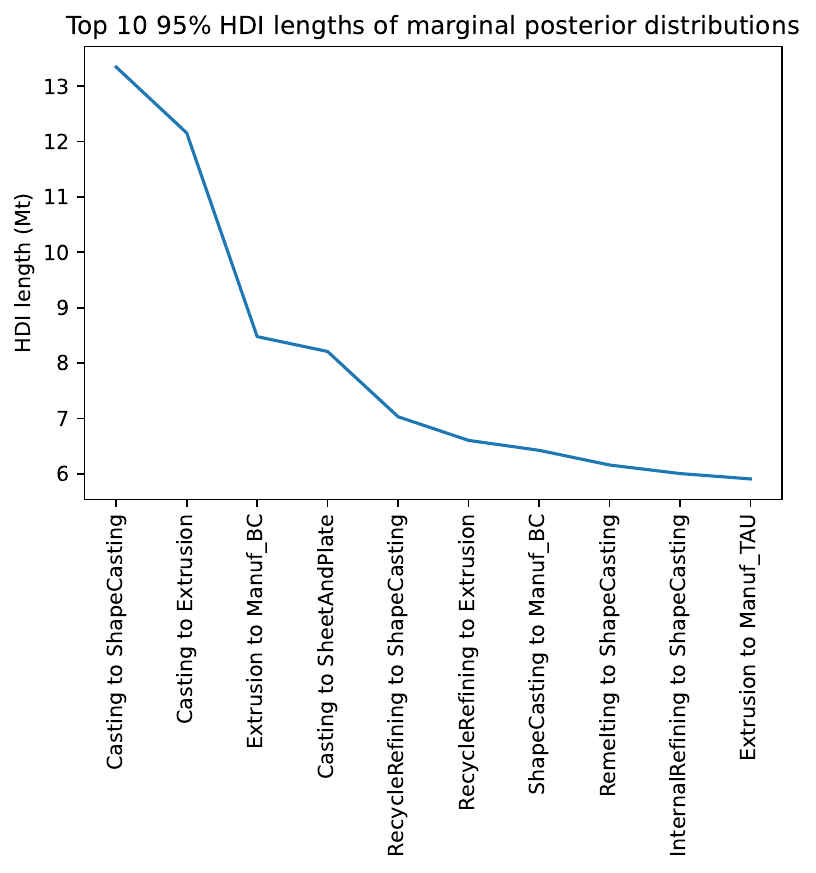}
\caption{}
\label{fig:top10largesthdiA}
\end{subfigure}
\begin{subfigure}[b]{0.48\textwidth}
\includegraphics[width = 1.00\textwidth,clip,trim = 0cm  0.0cm 0cm 0cm]{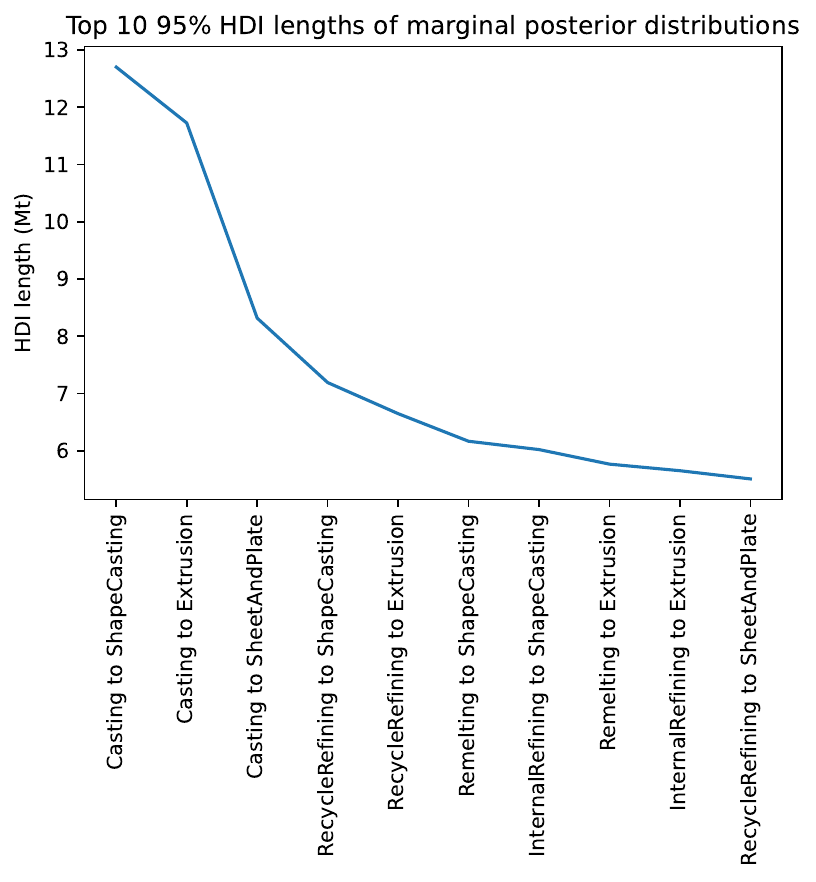} 
\caption{}
\label{fig:top10largesthdiB}
\end{subfigure}
}
\caption{Graphs of posterior predictive checks and posterior predictive HDI lengths, for scenario B. On the first row, we have the sample $95\%$ posterior predictive highest density intervals (red bar) and the observed values (blue dot) for the change in stock data, observed flow data, conservation of mass conditions and flow ratio data respectively. On the second row, we have the posterior predictive p-values for the observed change in stock variables, observed flow variables, conservation of mass conditions and flow ratio data respectively. On the third row, we plot the top $10$ largest marginal posterior HDI lengths for scenario A (left), and scenario B (right). Underlying data for this figure can be found at https://github.com/jwang727/BayesianMFA }
\label{fig: aluminiumposteriorpredplots}
\end{figure*}
  
In this section we present some additional results for the aluminium model. First, we apply the posterior predictive checks described in \Cref{sec: posteriorpredictivechecks} on the aluminium model under scenario B. From \Cref{fig:stockpredictiviehdi,fig:flowpredictiviehdi,fig:CoMpredictiviehdi,fig:ratiopredictiviehdi}, it can be seen that the $95\%$ HDIs all contain the observed values for all change in stock data, flow data, conservation of mass conditions and ratio data. Moreover, from \Cref{fig:stockpvalues,fig:flowpvalues,fig:CoMpvalues,fig:ratiopvalues}, it can be seen that the posterior predictive p-values are mostly between $0.3$ and $0.7$ and no p-value is smaller than $0.05$ or greater than $0.95$, suggesting that no p-values are extreme and the model generally fits the data reasonably well. There are a few variables close to extreme values however. In particular, the 27th flow data in \Cref{fig:flowpvalues} has a p value of around $0.07$, which represents the total outflow from `Internal remelting' to `Semi-manufacturing' of $9.3$ Mt. Upon reviewing the data, it was found the total inflow of `Internal remelting' only summed to $7.1$ Mt. The posterior predictive check therefore helpfully highlighted a discrepancy that suggests the data for the process Internal remelting has more sizable mass imbalance. Similarly, the 49th data has a relatively high p-values of around $0.8$, which is the data representing the outflow from Manuf\_OTD of $0.8$ Mt. Upon examining the data it was found the total inflow of Manuf\_OTD is $1.0$.
 
In MFA, data is expensive to collect, so it is useful to prioritise which flow and change in stock variables to collect more data on. This will likely depend on what questions the modeller is most interested in answering regarding the real system being modelled. Without specific questions however, Bayesian inference provides default strategies for prioritising which data points to collect, by ranking the variables in terms of descending posterior uncertainty. We plot the top $10$ flow and change in stock variables in descending length of their marginal $95\%$ HDI for both scenarios A (\Cref{fig:top10largesthdiA}) and B (\Cref{fig:top10largesthdiB}). In scenario B, the most uncertain variables are mostly flows from `Casting' or `Recycling' to `Semi-manufacturing', which is expected as those are the disaggregated flows where data is not available. For scenario A, the most uncertain variables are more scattered throughout the system, as scenario A has no data on any disaggregated flows.

\section{Discussion} \label{sec: discussion}

This paper presented a novel MFA methodology under the Bayesian framework that addresses existing challenges and expands the applicability of Bayesian inference in MFA. By relaxing the mass balance constraints with a noise term, we improved upon the computational scalability and reliability of posterior samples compared to existing methods, while still retaining well mass balanced posterior estimates of stock changes and flows. We introduced a child and parent parametrisation that can conveniently deal with MFA systems with multiple layers of disaggregation of processes and flows, providing posterior distributions on flows and stock changes on all levels of disaggregation in the system, including lower levels where data is often unavailable. We showed that even an weakly informative prior, specifically a prior based on the nearest order of magnitude of stock changes and flows, is capable of greatly improving the model's estimation accuracy and quality of its uncertainty quantification, especially during the early stages of the analysis when there is a lack of data, reaffirming the benefit of a Bayesian approach to MFA. We also demonstrated how posterior predictive checks can be used to check if the model is consistent with the data and mass balance conditions, and help identify data inconsistencies and aid in selecting noise parameter values for the data and mass balance conditions.

Like other Bayesian approaches, our method requires prior distributions of all flow and stock change variables of interest to be manually specified, which is an additional requirement compared to traditional MFA. However, we argue that this should be a standard part of any material flow analysis, where domain knowledge should be continuously collated, and the prior distribution offers a principled, mathematical way of incorporating this knowledge into the model. The priors used in this paper are unimodal and only weakly informative at most, in order to reduce requirements on the prior so that it can be applied in a wider range of MFA settings. In principle more or less informative priors can be used based on domain knowledge of the application. In the presence of multiple expert opinions, mixture priors that combine multiple experts similar to the approach of \cite{Dong2022} can be considered in our model framework.

The Bayesian approach inevitably comes with a higher computational cost, as the full posterior distribution of each variable of interest needs to be calculated rather than just a point estimate, and may run into convergence issues \cite{Betancourt2017}. The No-U-Turn Sampler HMC algorithm used for our model took around $90$ minutes to generate $24000$ posterior samples across two chains for the aluminium dataset, which we consider acceptable. We also note that most MFA literature do not use very large datasets so we do not anticipate computational cost to be a major issue for most MFA studies. However, for much larger material flow datasets, approximate methods such as variational Bayes can be employed to reduce the computational cost, or minimally just estimating the posterior mode directly (and forgoing uncertainty quantification) can potentially yield better point estimates than a non Bayesian method if informative priors are available.





\paragraph*{Acknowledgements}
This work was supported by the UKRI Interdisciplinary Circular Economy Centre For Mineral-based Construction Materials under the EPSRC Grant EP/V011820/1. The authors are grateful to Mohit Arora, Nicola Gambaro, Ugo Legendre and Shen Zhenxia for useful discussions. 


\bibliographystyle{abbrvnat}
\bibliography{bibliography}

\appendix
\newpage

\section{Supplementary Information}

\vspace{2cm}




\subsection{Parent and child processes}  \label{sec: parentandchildprocess}

\setcounter{MaxMatrixCols}{20}

In this section we elaborate on the parent and child process framework used to model disaggregation of processes or flows in MFA systems, and provide a simple illustrative example. Recall that a child process is any process which does not contain subprocesses, and a parent process is any process which does contain subprocesses. This by definition partitions all processes in the system into either child or parent processes, and all parent processes can be broken down into a set of constituent child processes. This partition of parent and child processes is useful because it allows multiple levels of disaggregation to be modelled using just two types of processes with a bottom up approach. In particular, only the child processes need to be modelled, and the parent processes variables are simply the sum of its constituent child processes variables. To illustrate this, consider the (non mass conserved) simple example of \Cref{fig: simplemfaexample}:

\begin{figure*}[h]{
\centering
\includegraphics[width = 1.00\textwidth]{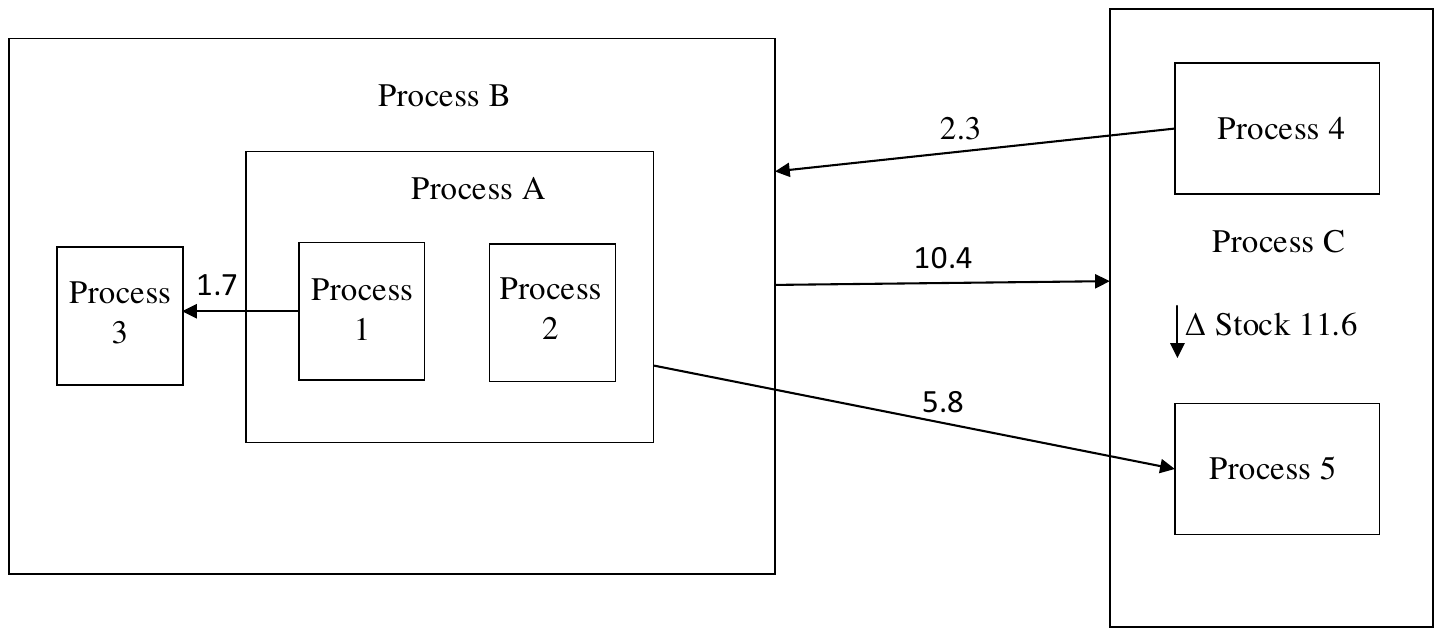}
}
\caption{A simple example of a MFA system with aggregation}
\label{fig: simplemfaexample}
\end{figure*}

From a typical top down approach, this example system contains 3 levels of disaggregation, since `Process B' contains `Process A', and `Process A' in turn contains `Process 1' and `Process 2'. In our framework however, the child processes are `Process 1',`Process 2',`Process 3',`Process 4' and `Process 5' while the parent processes are `Process A', `Process B' and `Process C'. The data in \Cref{fig: simplemfaexample} are the flows $U_{1,3}=1.7$, $U_{4,B}=2.3$, $U_{B,C}=10.4$ and $U_{A,5}=5.8$, and the change of stock $S_C=11.6$. The change in stock or flow data involving parent or aggregated processes can all be expressed in terms of flows involving child processes:

\begin{align} 
S_C&=S_4+S_5 \\
U_{4,B}&=U_{4,1}+U_{4,2}+U_{4,3} \\
U_{B,C}&=U_{1,4}+U_{2,4}+U_{3,4}+U_{1,5}+U_{2,5}+U_{3,5} \\
U_{A,5}&=U_{1,5}+U_{2,5} 
\end{align} 
The data in this example can therefore be formulated with the following design matrix:

\begin{equation} 
\begin{pmatrix}
0 & 0 & 1 & 0 & 0 & 0 & 0 & 0 & 0 & 0 & 0 & 0 \\
1 & 1 & 0 & 0 & 0 & 0 & 0 & 0 & 0 & 0 & 0 & 0 \\
0 & 0 & 0 & 0 & 0 & 0 & 0 & 0 & 0 & 1 & 1 & 1 \\
0 & 0 & 0 & 1 & 1 & 1 & 1 & 1 & 1 & 0 & 0 & 0 \\
0 & 0 & 0 & 0 & 1 & 0 & 1 & 0 & 0 & 0 & 0 & 0
\end{pmatrix}
\begin{pmatrix}
S_4 \\
S_5 \\
U_{1,3} \\
U_{1,4} \\
U_{1,5} \\
U_{2,4} \\
U_{2,5} \\
U_{3,4} \\
U_{3,5} \\
U_{4,1} \\
U_{4,2} \\
U_{4,3}
\end{pmatrix}
=
\begin{pmatrix}
1.7 \\
11.6 \\
2.3 \\
10.4 \\
5.8 
\end{pmatrix}
\end{equation} 

\newpage

\subsection{Model convergence diagnostics} \label{sec: modeldiagnostics}

\begin{figure*}[h]{
\centering
\includegraphics[width = 1.00\textwidth]{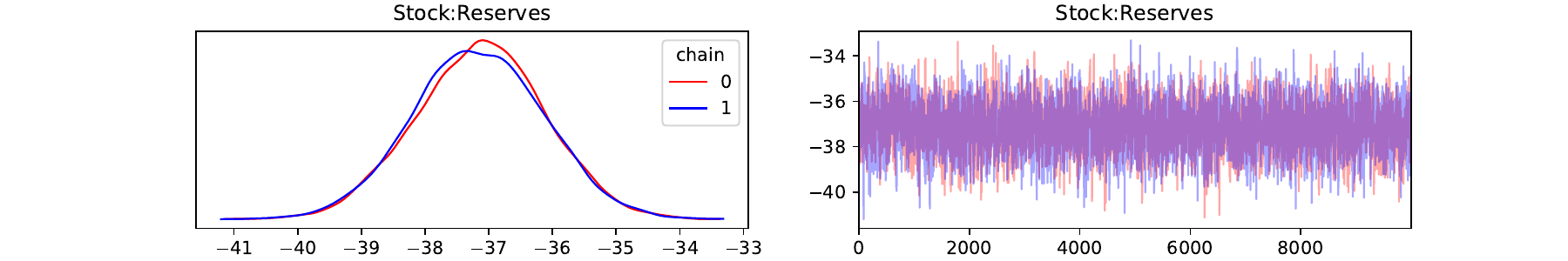}
\includegraphics[width = 1.00\textwidth]{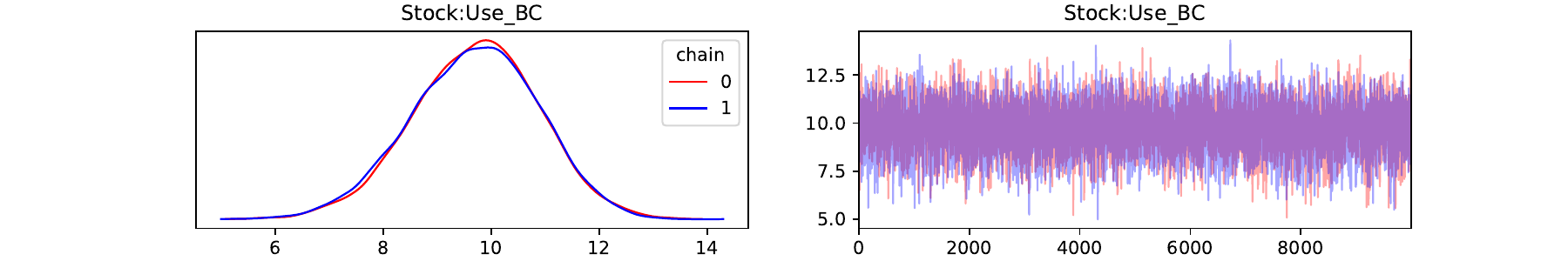}
\includegraphics[width = 1.00\textwidth]{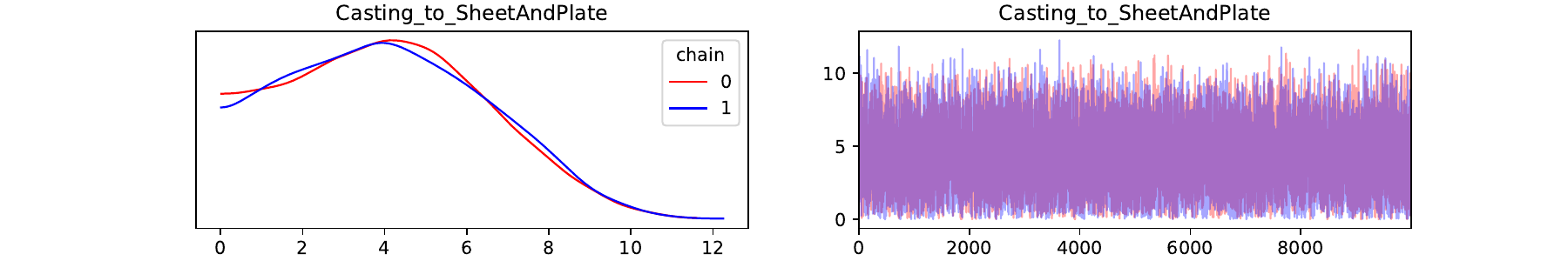}
\includegraphics[width = 1.00\textwidth]{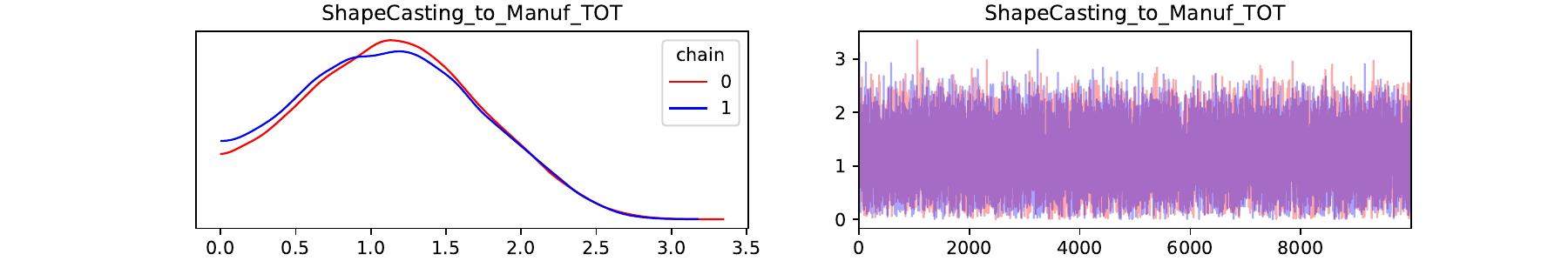}
\includegraphics[width = 1.00\textwidth]{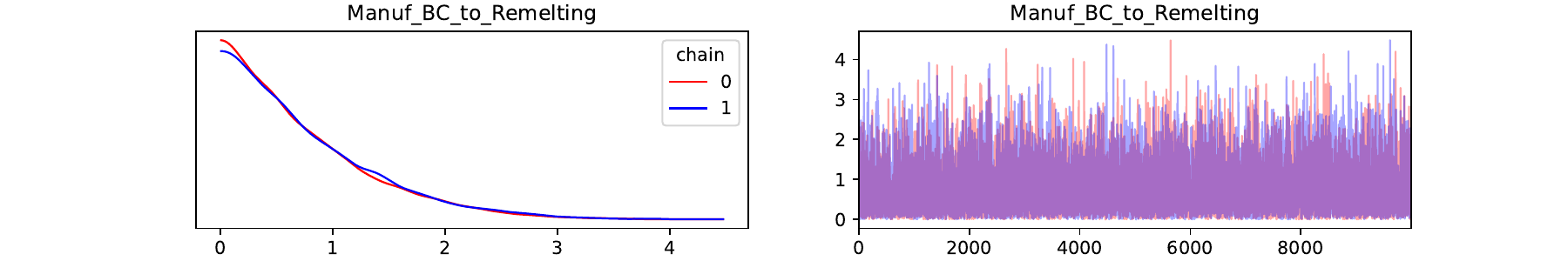}
\includegraphics[width = 1.00\textwidth]{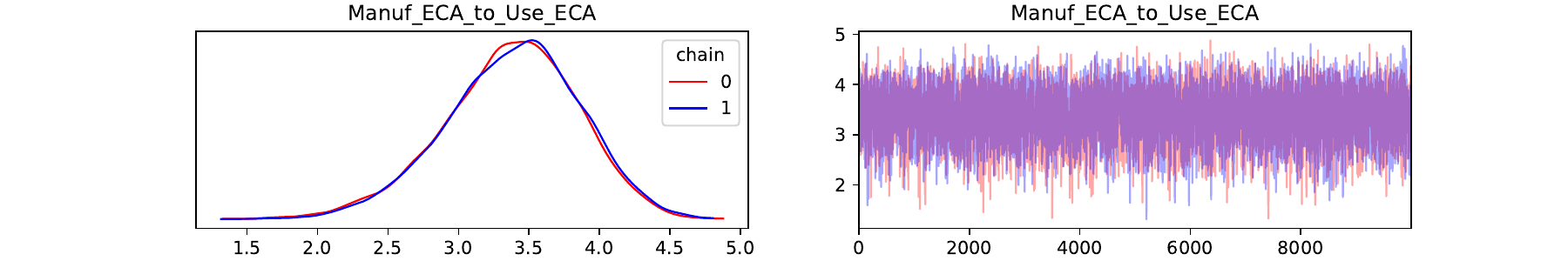}
}
\caption{Traceplots for a selection of flow and change in stock variables in the aluminium model, scenario A.}
\label{fig: aluminiumtraceplots}
\end{figure*}

In this section we describe model diagnostic checks used to examine the convergence of the Markov Chains Monte Carlo (MCMC) sampling algorithm (No-U Turn Sampler of \cite{Hoffman2014}). We present trace plots for a selection of flow and change in stock variables in the aluminium model for illustration. Using the No-U-Turn Sampler, we sampled 2 independent chains of $12000$ samples, with the first $2000$ samples used for tuning and discarded from the final posterior samples. The purpose of the tuning samples are to give the Markov Chains iterations to converge to the target posterior distribution before extracting samples to approximate the posterior distribution. The traceplots in \Cref{fig: aluminiumtraceplots} suggests both chains have converged to the same distribution. Furthermore, for both scenario A and scenario B of the aluminium model, the Gelman-Rubin diagnostic statistic (see for example Chapter 11 of \cite{Gelman2013}) for all flow and change in stock variables in the model have converged to $1.0$ and no divergent samples were reported, indicating the posterior samples are reliable (\cite{Betancourt2017}). 

\newpage

\newpage

\subsection{Posterior pairplots} \label{sec: posteriorpairplots}

Here we present a small selection of pairplots to illustrate correlation between the posterior distribution of flow variables. In particular, we display all the outflows of the `Foil' process, which is a subprocess of semi-manufacturing. The outflow processes of `Foil' are `InternalRemelting', `Manuf\_TAU', `Manuf\_TOT', `Manuf\_POT', `Manuf\_CD'. Here one can see from the pairplots that it's very uncommon for the total of any two outflows to exceed around $6$ Mt, which is consistent with the data which has the total outflow of `Foil' at $6.4$ Mt.

\begin{figure*}[h]{
\centering
\includegraphics[width = 1.00\textwidth]{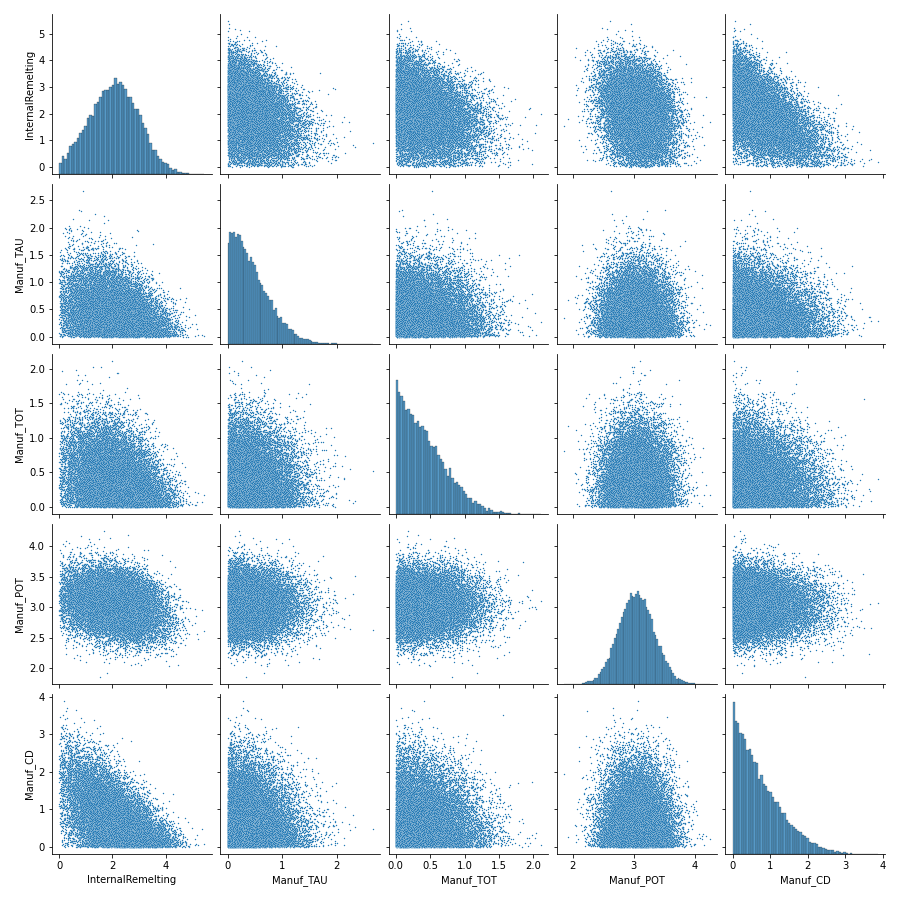}
}
\caption{Pairplots for the outflows of the Foil process, scenario A. Axis units are in Mt. }
\label{fig: aluminiumpairplots}
\end{figure*}

\newpage

\subsection{Detail of prior parameters for the aluminium model} \label{sec: priorparametersAl}

In this section we give detail on the choice of prior parameters for the model evaluation of the aluminium dataset, A summary of the parameter values described in this section can be found in \Cref{table: aluminiumparameters}. Below we describe and motivate the choices:

The prior mode $\mu_i$ or $\mu_{j,k}$ of flow and change in stock variables respectively are chosen to be equal to the nearest power of $10$ of the reported value when it is available. For example, the flow from `Mining' to `Refining' with a reported value $35.4$ Mt is assigned a prior mode of $\mu_{1,2}=10.0$ Mt. If the flow or change in stock variable has no reported value, it is assigned an uninformative prior mode of $1.0$ Mt instead.  

The prior standard deviation $\sigma_i$ for the change in stock variables are chosen to be $\sigma_i=\max(\sqrt{40}|\mu_i|,0.1)$, or $10.0$ if there is no reported value for the flow or change in stock variable. The corresponding parameter $\sigma_{j,k}$ for the flow variables are chosen similarly. For example, $\sigma_{1,2}=\sqrt{40}*10.0=63.25$ for the flow from `Mining' to `Refining'. This choice is motivated by when $\mu_i$ is specified to the correct order of magnitude, the true value of the variable is at most $4.5|\mu_{i}|$ away, and so $\sigma_i$ is chosen on a similar scale but somewhat larger to ensure the prior is not over-confident. The maximum here is to ensure a minimum prior standard deviation of $0.1$. 

Note we are not advocating for this choice of prior distribution to be uniquely correct. The overall rationale of this choice for the prior is to assign a weakly informative prior distribution with a prior mode that captures the order of magnitude of the flow or change in stock, coupled with a relatively large prior variance, in order to reflect a realistic baseline of domain knowledge to aim for when conducting MFA. 

For lack of better information, the standard deviation parameters for change in stock data noise $\tau_{i'}$ are chosen to be $\tau_{i'}=\max(0.1|Y_{i'}|,0.1)$, where $Y_{i'}$ are the observed change in stock data. The corresponding parameters $\tau_{j'}$ for flow data noise is chosen similarly. In other words, the data standard deviation is equal to $10\%$ of the observed data as in \cite{Lupton2018}, but with a minimum of $0.1$ to account for a minimum degree of uncertainty from rounding or other measurement errors. For the ratio data noise, the standard deviation parameters  $\tau_{k'}$ are chosen with a somewhat smaller minimum $\tau_{k'}=\max(0.1Y_{k'},0.01)$, where $Y_{k'}$ is the observed ratio. While the choice of standard deviation parameters here is somewhat arbitrary, the posterior predictive checks performed on the aluminium model do not exhibit extreme Bayesian p-values (smaller than 0.05 or greater than 0.95), which suggests the standard deviation parameters chosen are reasonable. This also suggests one way to tune standard deviation parameters is to start with a small choice of standard deviation parameters (such as $10\%$ of the observed data value), run the model, identify the data points which exhibit extreme Bayesian p-values in the posterior predictive checks, and increase the standard deviation parameter for those points and rerun the model until no extreme Bayesian p-values remain.   

For the mass conservation conditions, a small constant standard deviation of $\tau_{l'}=0.05$ is chosen for all mass balance conditions. Ideally these parameters should be chosen based on the modeller's confidence regarding epistemic uncertainty in the system, giving higher standard deviations to processes where there is more uncertainty in the system definition. Like with the standard deviation parameters for the data, posterior predictive checks can help select suitable parameter values, where the standard deviation parameters for mass balance conditions with extreme Bayesian p-values should be increased until the Bayesian p-values are no longer extreme. Without prior confidence of epistemic uncertainty in the system, we recommend starting with a small constant standard deviation for the mass conservation conditions, but not too small so that the mass conservation conditions are sufficiently relaxed to ensure the NUTS algorithm converges well. In this case it turned out the choice of $\tau_{l'}=0.05$ did not produce any extreme Bayesian p-values, produces well converged posterior samples and well mass balanced flows (see subsequent section for more detail). 

\newpage

\begin{table}[h]
\centering
\noindent \begin{tabular}{| p{2.0cm}| p{4.8cm}|}
 \hline
 Parameter & Parameter prior value \\
 \hline
 $\mu_i$ & nearest power of $10$ of reported value, or $1.0$ if unavailable  \\
   \hline
 $\mu_{j,k}$ & nearest power of $10$ of reported value, or $1.0$ if unavailable  \\
 	\hline
 $\sigma_i$ & $\max(\sqrt{40}|\mu_i|,0.1)$, or $10.0$ if reported value is unavailable  \\
 \hline
 $\sigma_{j,k}$ & $\max(\sqrt{40}\mu_{j,k},0.1)$, or $10.0$ if reported value is unavailable  \\
 \hline
 $\tau_{i'}$ & $\max(0.1|Y_{i'}|,0.1)$ \\
 \hline
 $\tau_{j'}$ & $\max(0.1Y_{j'},0.1)$ \\
 \hline
 $\tau_{k'}$ & $\max(0.1Y_{k'},0.01)$ \\
 \hline
 $\tau_{l'}$ & 0.05 \\
 \hline
  \end{tabular} 
  \caption{Table of parameters for the aluminium model}
\label{table: aluminiumparameters}
\end{table}

\newpage

\subsection{Assessment of posterior mass balance for the aluminium model} \label{sec: massbalanceassess}

In this section we analyse how well the mass balance conditions are satisfied by the aluminium model, given the parameters chosen in the previous section, in particular the standard deviation parameters $\tau_{l'}=0.05$ for the mass balance conditions. We examine the posterior distribution of the mass balance conditions to see how well they are centred around $0$.

\begin{figure*}[h]{
\centering
\includegraphics[width = 0.50\textwidth]{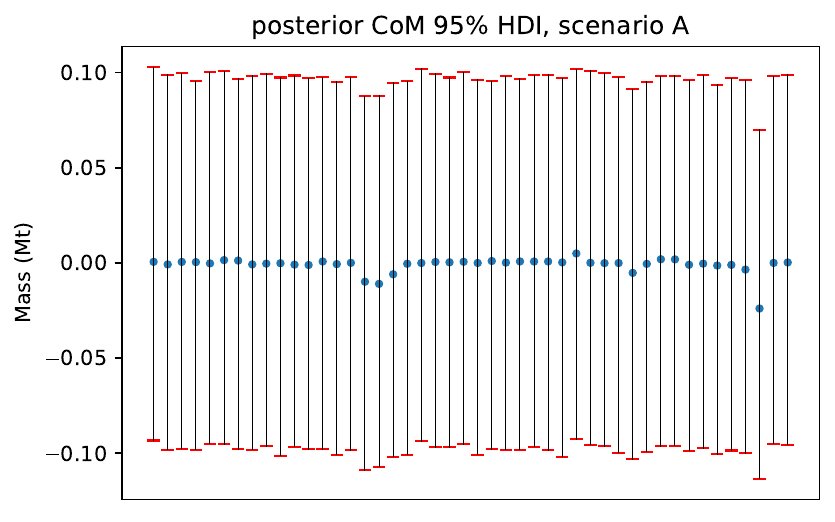}
\includegraphics[width = 0.50\textwidth]{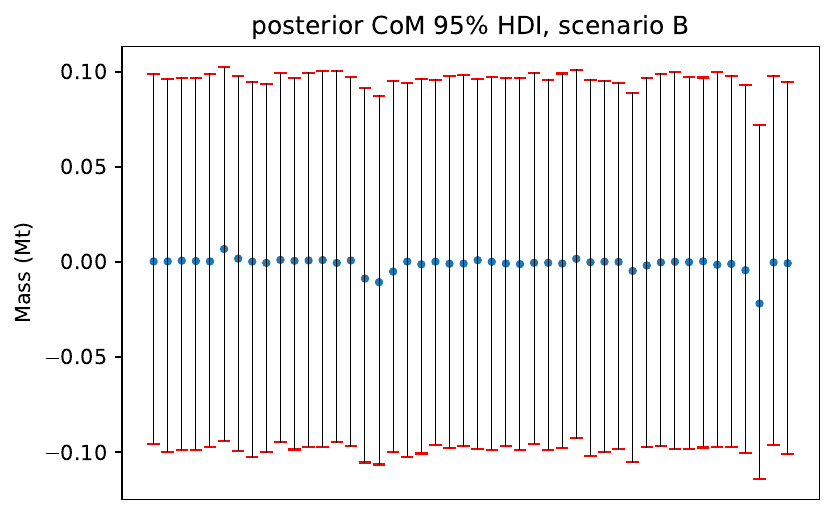}
\caption{posterior HDI lengths for the mass conservation conditions. On the top we have scenario A and on the bottom we have scenario B. The $95\%$ posterior highest density intervals are shown with a red bar and the posterior mean is shown with a blue dot}
}
\label{fig: posteriormassconserve}
\end{figure*}


From the Figure above, it can be seen that the posterior mean of the mass balance conditions is very close to $0$ in both cases, having a mean of $-0.001$ Mt for both scenario A and scenario B when averaged over all processes, which is orders of magnitudes smaller than the average posterior flow magnitude of around $2.3$ Mt for both scenario A and B. This ensures that the posterior mean of the flow and stock change variables in the system, which is also the point estimate we report, is very well balanced. The individual posterior samples are somewhat less mass balanced, exhibiting a standard deviation of around $0.05$ Mt for all processes. However, this is still a small level of mass imbalance and it is not necessary for every posterior sample to be perfectly mass balanced for the model to produce useful point estimation or uncertainty quantification. 

\newpage

\subsection{Zinc cycle simulation study} \label{sec: zinccasestudy}

\begin{figure}[h]
\includegraphics[width = \textwidth]{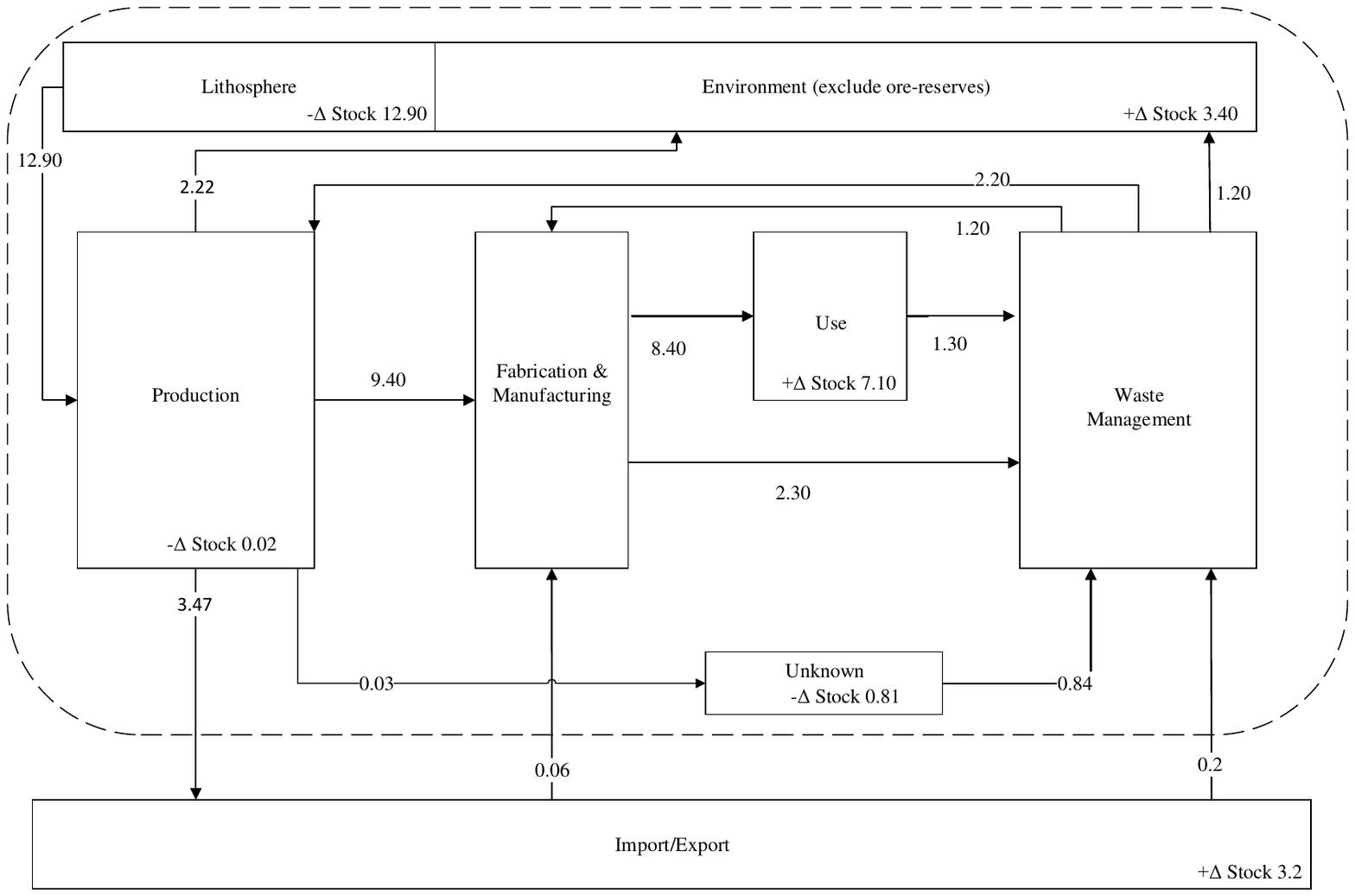}
\caption{Zinc cycle for China, ca. 1994-1998, adapted from Figure 2d of \cite{Graedel2008}. The unit of mass for zinc is displayed in $10^5$ metric tons per year. we explicitly define an `Unknown' Process to account for flows with an unknown source or destination.
}
\label{fig: zincflowdiagram}
\end{figure}

We conduct a simulation study to investigate the estimation accuracy and uncertainty quantification of our model. For this end we consider the zinc cycle in China, ca. 1994-1998, from \cite{Graedel2008}. The zinc cycle contains no missing data values, so for the purposes of this simulation, we treat the reported data values as the true parameter values of the flow and stock change variables. This allows us to assess our model by withholding some data from it and checking the model output against the true values of the stock change and flows. We examine how our model performs in terms of estimation accuracy and uncertainty quantification under different levels of prior knowledge, as data is incrementally added into the model. The zinc model considered is linear for ease of comparison with other models that require data to be in the form of a design matrix. However, we note that common MFA data types including flows, flow ratios and stock changes can be expressed under a linear model.

To assess estimation accuracy of the model, we calculate how the error of the model evolves as data is randomly added into the model one at a time, starting from no data, until the full dataset becomes available to the model. We consider two types of error, the root mean square error (rmse) and the maximum error, of the model's predictions compared to the reported values of the flow and change in stocks variables. The posterior mode is taken as the model prediction. To account for the random order in which data is added into the model, we repeat this process 50 times and calculate the average error (for both rmse and the maximum error) over the 50 runs for each possible data size from $0$ to $20$. 

For comparison, we consider several alternate models. First, we consider an alternate Bayesian Gaussian model which has the advantage of a closed form solution, but carries the disadvantage of assigning significant probability to negative values for flow variables in the posterior distribution, which should be nonnegative as negative flows are not physically meaningful. Detail of the Gaussian model is provided in \Cref{sec: gaussianmodel}. In addition, we also investigate the effect of different levels of prior knowledge has on the error. We utilise two different levels of prior knowledge, the first is an `uninformative' prior which sets the prior mode of all variables to the average absolute value of the full data, and the correct sign for change in stock variables. This intuitively represents a prior that knows the average order of magnitude of the system but has little visibility on the individual magnitude of each flow or change in stock. The second is a `weakly informative' prior which sets the prior mode of each flow and change in stock variable to the nearest power of $10$ of the true value. This intuitively represents a prior that has knowledge of each flow and change in stock variable to the nearest order of magnitude. We also provide results from non Bayesian regression models for comparison, specifically ridge regression and multilayer perceptron (see e.g. \cite{ESLII}). For each run, both algorithms are trained on the same datasets as the proposed and Gaussian model, and outputs a prediction on the vector of all flows and change in stock variables. While these models do not have a principled way of incorporating prior information, one can nevertheless train those models on the data and use them to estimate the value of all flow and change in stock variables in a system, which makes a comparison with Bayesian MFA models meaningful.

\begin{figure}[t!]
\centering
\includegraphics[width = 0.48\textwidth]{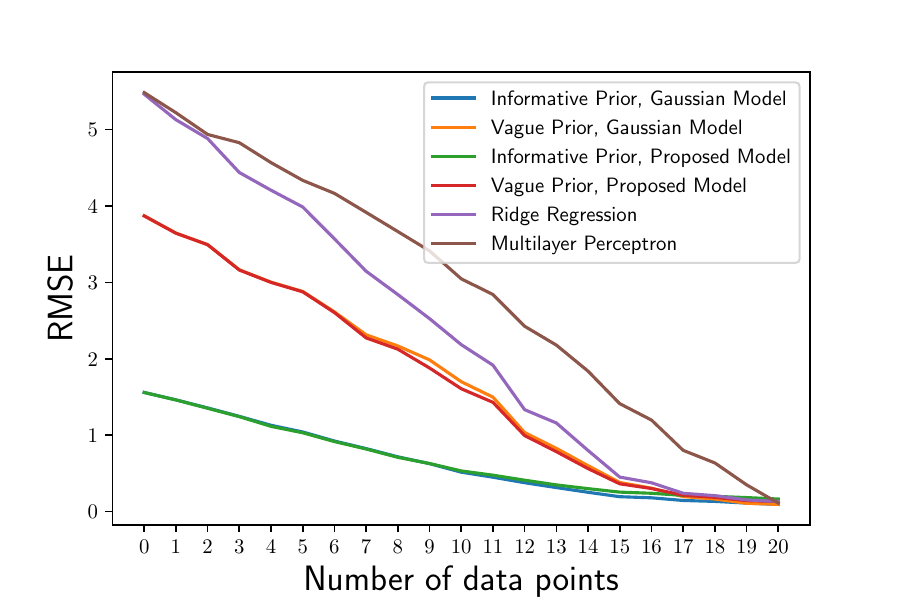}
\includegraphics[width = 0.48\textwidth]{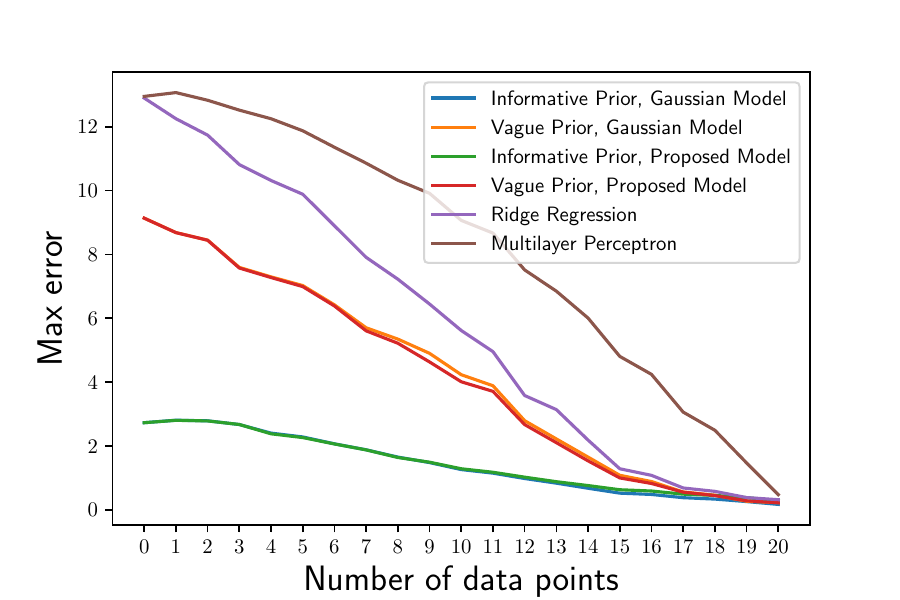}
\caption{Zinc flow data error analysis:
(left) Root mean squared error RMSE of the model plotted against number of data points available to the model, averaged over 50 runs or datasets.
(right) Maximum error of the model plotted against the number of data points available to the model, averaged over 50 runs or datasets. For each error type, our model and the Gaussian model are considered under uninformative and informative priors. Ridge regression is provided as a non Bayesian baseline. 
}
\label{fig: errorzinc}
\end{figure}

\Cref{fig: errorzinc} shows the results of our analysis, where the rmse and maximum error of the model are plotted against the number of data points available to the model, averaged over 50 runs. For ridge regression and multilayer perceptron, the starting rmse at no data is around 5.4, whereas for the uninformative prior it is around 4.0 and for the weakly informative prior around 1.5. As data is added, all models exhibit the trend of monotonically decreasing error, but the proposed and Gaussian models exhibit noticeably lower error until almost all the data becomes available to the model, indicating incorporating prior information leads to more accurate estimates. Notably, it takes around 10 data points for the uninformative prior models and 12 data points for ridge regression to achieve the starting rsme (around 1.5) of the weakly informative prior model. Therefore, one can consider only knowing the order of magnitude of each flow or change in stock to be roughly equivalent to $10$ data points in terms of rmse in this model, demonstrating the utility of being able to incorporate knowledge via a Bayesian prior in such an underdetermined system. For the Gaussian model, theoretical bounds on the mean square error provided in \Cref{sec: gaussianmodel} also support the trend exhibited in \Cref{fig: errorzinc}, and the intuition that having an informative prior can greatly increase the model's accuracy for MFA problems when there is a shortage of data but available expert domain knowledge.

To assess uncertainty quantification of the posterior distribution, we examine whether the $95\%$ posterior marginal highest density intervals (HDIs) of our model have similar level of frequentist coverage, i.e. form correct confidence intervals. Unlike credible intervals which have a designated posterior probability of containing the unknown parameter, confidence intervals are constructed such that upon repeated experiments (i.e. generate samples of data), a designated proportion (known as coverage) of the confidence intervals constructed will contain the true value of the unknown parameter. Bayesian credible intervals are in general not guaranteed to have good coverage, particularly in high-dimensional settings that are typical in MFA, but nevertheless it is desirable for them to have good coverage. Coverage is a property of the model procedure and does not require the repeated experiments to be the same experiment or dataset. It is therefore desirable for uncertainty quantification generated by a model procedure to contain the true value of the unknown parameter most of the time, over many runs for possibly different datasets. We investigate the coverage of our model's credible intervals empirically in order to understand whether they can be used reliably as confidence sets.

The high dimensional nature of MFA models means the dimension of parameter space $p$ (corresponding to the number of flow and change in stock variables in the model) is usually greater or similar in size compared to the number of available data $n$, and $n$ is typically not large in magnitude. Therefore, results which guarantees asymptotic frequentist coverage of credible regions, such as the Bernstein-von Mises theorem (see e.g. \cite{Vaart1998}), may not apply in MFA problems. We therefore study in simulations the frequentist coverage of our model's posterior credible intervals, in an effort to understand whether these can be used as reliable confidence intervals. Specifically, we sample $300$ sets of the zinc dataset $\bm{Y}_1,\bm{Y}_2,\dots,\bm{Y}_{300}$ from the model likelihood $p(\bm{Y}|\bm{\theta}_0)$, where we take the `true values' $\bm{\theta}_0$ to be the originally reported flow and change in stock values reported in \cite{Graedel2008} . For each dataset sampled, we add data to the model in batches of $5$ datapoints at a time (in the order listed from top to bottom in \Cref{table:zinccoveragetable}), and calculate the posterior marginal HDIs to see if they contain the true value for each dataset size $n=5,10,15,20$. The frequencies at which the posterior marginal HDIs contain the true value over the $300$ runs, known as the coverage probability, are reported in \Cref{table:zinccoveragetable}. 

\begin{table}[h!]
\centering
\scriptsize
\noindent \begin{tabular}{ |p{2.6cm}| p{1.0cm}|p{1.0cm}|p{1.0cm}|p{1.0cm}| p{1.0cm}|p{1.0cm}|p{1.0cm}|p{1.0cm}| }
\hline
  &   \multicolumn{8}{c|}{Coverage probability(\%), average width of $95\%$ HDI }   \\
\hline
  &   \multicolumn{4}{c|}{Weakly informative prior}   &   \multicolumn{4}{c|}{Uninformative prior} \\
 \hline
 Variable Name & 5 data & 10 data & 15 data & 20 data & 5 data & 10 data & 15 data & 20 data\\
 \hline
 $\Delta$Stock: Use   & 95.3, 3.60 & 96.3, 3.44 & 98.0, 3.16 & 96.3, 3.15 & 96.0, 3.76 & 95.3, 3.63 & 97.7, 3.23 & 95.7, 3.21  \\
   \hline
 Production to ImportExport   & 90.7, 3.56 & 96.3, 3.40 & 96.7, 3.41 & 98.0, 2.92 & 91.3, 3.78 & 95.7, 3.65 & 97.7, 3.60 & 96.0, 3.15  \\
 	\hline
 $\Delta$Stock: Environment   & 98.0, 3.50 & 97.0, 3.35 & 97.3, 3.35 & 98.7, 3.18 & 95.0, 3.66 & 97.3, 3.44 & 96.0, 3.46 & 98.3, 3.25  \\
 	\hline
 WM to Production & 94.7, 3.39 & 96.0, 3.15 & 96.7, 3.12 & 93.0, 3.02 & 94.7, 3.54 & 94.0, 3.47 & 94.7, 3.35 & 91.3, 3.24  \\
 	\hline
 ImportExport to WM & 100.0, 0.83 & 100.0, 0.82 & 100.0, 0.82 & 100.0, 0.79 & 96.0, 2.29 & 99.3, 2.16 & 97.0, 2.12 & 98.7, 1.82  \\
 	\hline
 Unknown to WM & 100.0, 5.07 & 99.0, 2.48 & 99.3, 2.18 & 100.0, 2.19 & 100.0, 8.52 & 99.0, 2.53 & 98.0, 2.42 & 97.7, 2.48 \\
    \hline
 Lithosphere to Production & 100.0, 7.73 & 96.3, 3.37 & 95.3, 3.38 & 93.7, 2.87 & 0.00, 11.3 & 96.0, 3.63 & 94.7, 3.55 & 94.7, 2.97  \\
 	\hline
 $\Delta$Stock: Production   & 100.0, 0.39 & 100.0, 0.39 & 100.0, 0.39 & 100.0, 0.39 & 100.0, 16.1 & 95.0, 3.67 & 95.3, 3.59 & 96.0, 3.50   \\
  \hline
 Production to F\&M   & 100.0, 7.69 & 94.7, 3.32 & 99.0, 3.19 & 93.7, 3.06 & 3.33, 8.25 & 89.0, 3.55 & 91.7, 3.39 & 87.0, 3.21 \\
 	\hline
 WM to Environment & 100.0, 4.19 & 98.0, 2.69 & 98.7, 2.50 & 99.0, 2.55 & 100.0, 4.88 & 96.3, 2.98 & 97.3, 2.72 & 95.7, 2.72  \\
 	\hline
 F\&M to Use   & 100.0, 7.54 & 99.7, 7.20 & 96.3, 3.03 & 93.0, 2.95 & 100.0, 9.55 & 100.0, 9.57 & 96.3, 3.14 & 93.3, 3.02  \\
 	\hline
 $\Delta$Stock: Unknown   & 100.0, 6.58 & 100.0, 4.62 & 100.0, 2.97 & 98.7, 2.97 & 100.0, 13.3 & 100.0, 7.20 & 98.0, 3.16 & 99.3, 3.16  \\
 	\hline
 Production to Unknown   & 100.0, 0.20 & 100.0, 0.20 & 100.0, 0.20 & 100.0, 0.20 & 100.0, 9.31 & 97.7, 4.90 & 93.7, 2.03 & 96.7, 2.01  \\
 	\hline
 F\&M to WM   & 100.0, 6.44 & 100.0, 6.52 & 93.3, 3.30 & 94.3, 3.12 & 100.0, 10.6 & 100.0, 11.7 & 93.3, 3.45 & 93.7, 3.21  \\
 	\hline
 Use to WM   & 100.0, 5.91 & 100.0, 5.93 & 98.3, 2.60 & 98.0, 2.54 & 100.0, 7.32 & 100.0, 7.71 & 98.0, 2.63 & 98.7, 2.57  \\
 	\hline
 WM to F\&M & 100.0, 7.79 & 100.0, 7.29 & 100.0, 3.79 & 98.3, 2.63 & 95.0, 14.1 & 100.0, 12.4 & 100.0, 4.16 & 97.7, 2.74  \\
 	\hline
 $\Delta$Stock: ImportExport   & 97.0, 5.21 & 99.3, 5.10 & 99.7, 5.10 & 96.0, 3.09 & 84.0, 9.63 & 90.3, 7.30 & 95.3, 6.55 & 93.0, 3.26  \\
 	\hline
 $\Delta$Stock: Lithosphere   & 100.0, 8.21 & 99.0, 4.95 & 99.3, 4.95 & 95.3, 3.04 & 0.00, 12.2 & 99.0, 5.24 & 98.7, 5.18 & 95.7, 3.11 \\
 	\hline
 ImportExport to F\&M & 100.0, 0.86 & 100.0, 0.85 & 100.0, 0.82 & 100.0, 0.79 & 100.0, 7.19 & 100.0, 4.39 & 100.0, 3.17 & 98.0, 1.80  \\
 	\hline
 Production to Environment   & 100.0, 3.77 & 100.0, 3.74 & 99.3, 3.92 & 95.0, 2.90 & 100.0, 4.20 & 99.0, 3.74 & 99.7, 4.14 & 96.0, 2.97  \\
 	\hline
  \end{tabular} 
  \caption{Table of coverage probability and average width of 95\% highest density intervals (HDI) of the marginal posterior distributions of all flow  and change in stock variables in the zinc data under different amounts of data and prior information, computed over $300$ runs. Data was added in batches of $5$ to the model in the order that they are listed in this table (from top to bottom).}
\label{table:zinccoveragetable}
\end{table}

From \Cref{table:zinccoveragetable}, it can be seen that the 95\% HDI possess similar levels of coverage probability in most cases. The only notable exception being the uninformative prior at $5$ data points, where the coverage probability are close to $0$ for the flow variables `Lithosphere to Production' and `Production to F\& M', and the change in stock variable for `Lithosphere'. This is likely caused by the first $5$ data points not containing any data on the `Lithosphere' process, as well as containing little data on `Production', so the model was unable deduce the true magnitude of how much material was taken from the `Lithosphere' with an uninformative prior. On the other hand, the weakly informative prior does not possess this issue as the prior has sufficient coverage over the true value. In addition, the average width of the $95\%$ HDI are generally lower for the weakly informative prior until data for that variable becomes available, which is expected from a more informative prior.

\newpage

\subsection{Detail of parameters for the zinc models} \label{sec: priorparameterszinc}

In this section we give detail on the choice of model parameters for all models used for the zinc case study, including how the prior parameters were chosen. Recall that we considered two different priors for the zinc model, an uninformative prior and a weakly informative prior. For the uninformative prior, the prior modes $\mu_i$ of the change in stock variables are chosen to be $3.6575$ (measured in $10^5$ tonnes/Yr), equal to the mean of the absolute value of all the stock change and flow variables in the data, to represent a prior that only has a rough knowledge of the average order of magnitude of the system, but not the order of magnitude of each individual flow or change in stock. For weakly informative prior, $\mu_i$ are chosen the same way as the aluminium case, where the prior mode is set to the nearest power of $10$ of the `true' reported value. The prior standard deviation parameters $\sigma_i$ are chosen to be a constant $\sqrt{40}$ in the uninformative case, and $\sigma_i=\max(\min(4|\mu_i|,4),0.1)$ in the weakly informative case. Like in the aluminium model, $\sigma_i$ for the weakly informative prior is chosen at a similar scale compared to $4.5|\mu_i|$ but somewhat smaller and capped at a maximum of $4.0$ to reflect the fact that the size of flows are relatively smaller in the zinc data (when measured in $10^5$ tonnes/Yr, compared to the aluminium data which is measured in Mt). The corresponding prior parameters for the flow variables $\mu_{j,k}$ and $\sigma_{j,k}$ are chosen similarly.
 
The noise standard deviation parameters of the observed data is chosen to be a constant $\tau=1.0$ throughout for all data, and similarly for the mass conservation conditions. Here the noise parameters are chosen to be somewhat larger than the aluminium model to speed up the computation of the No-U-Turn Sampler MCMC algorithm, as the model needed to be computed $2400$ times over the varying degrees of prior knowledge and available data to generate the coverage probability results in Table 1 of the main text. In particular, small values of the noise parameter induces high curvature which makes the sampling algorithm take longer to explore the posterior distribution. Similarly, for assessing point estimation (root mean squared error and maximum error) of the model on the zinc data, the posterior mode of the model is estimated directly via optimisation for faster computational speed, rather than through approximating the entire posterior distribution using MCMC.  

\begin{table}[h]
\centering
\noindent \begin{tabular}{| p{2.0cm}| p{4.8cm}|p{4.8cm}|}
 \hline
 Parameter & Parameter prior value (weakly informative prior) & Parameter prior value (uninformative prior) \\
 \hline
 $\mu_i$ & nearest power of $10$ of reported value  & $ \pm 3.6575$ (sign depending if the change in stock is positive or negative) \\ 
   \hline
 $\mu_{j,k}$ & nearest power of $10$ of reported value  & $3.6575$ \\
 	\hline
 $\sigma_i$ & $\max(\min(4|\mu_i|,4),0.1)$ & $\sqrt{40}$ \\ 
 \hline
 $\sigma_{j,k}$ & $\max(\min(4\mu_{j,k},4),0.1)$ & $\sqrt{40}$ \\ 
 \hline
 $\tau$ & 1.0 & 1.0 \\
 \hline
  \end{tabular} 
  \caption{Table of parameters for the zinc model}
\label{table:zincparameters}
\end{table}

The ridge regression model was trained with a regularization parameter of $\lambda=1.0/40.0=0.025$, which was chosen to match the prior variance and noise standard deviation in the uninformative prior case for the Gaussian model. To see this, note the solution of the ridge regression is equal to the posterior mode of a model with prior $\bm{\theta} \sim \mathcal{N}(0, \sigma^2 I)$ and likelihood $\bm{Y}|\bm{\theta} \sim \mathcal{N}(X\bm{\theta}, \tau^2 I)$, and the regularization parameter can be shown to be equal to $\tau^2/\sigma^2$. So effectively the ridge regression solution is similar to the uninformative prior case, except the prior mean is set to $0$ instead which is arguably less informative. 

The multilayer perceptron regressor was trained with the default settings in the sklearn Python library, except with the maximum number of iterations increased to $1000$. In particular the default loss function is the squared loss and default number of hidden layers equal to $100$. 

\clearpage

\subsection{Gaussian model} \label{sec: gaussianmodel}

For model comparison purposes and to gain intuition into how Bayesian models perform for MFA problems, it is useful to examine a model with simplified assumptions. In particular, we consider a Gaussian model, where both the prior and the likelihood function are assumed to be normally distributed, and the data is assumed to be linear in terms of $\bm{\theta}$ (recall from the Methods section in the main text of the paper that the most common forms of MFA data can be expressed linearly in $\bm{\theta}$):

\begin{align}
\bm{\theta} & \sim \mathcal{N}(\bm{\mu}, \Sigma) \\
Y_i|\bm{\theta} & \sim \mathcal{N}(\bm{x_i}^\top\bm{\theta}, \tau_i^2) 
\end{align} \label{eq: gaussianmodel}
where $\bm{\mu}$ is the prior mean and $\Sigma$ is the (positive definite) prior covariance matrix. This gives a mathematically convenient \textit{conjugate} posterior that is also normally distributed, with its posterior mean $\bm{\mu}^{n}$ and covariance $\Sigma^{n}$ available in closed form:

\begin{align}
\bm{\mu}^{n} &= \bm{\mu}+\Sigma X^\top (X\Sigma X^\top+ T)^{-1} (\bm{Y}-X\bm{\mu}) \\
\Sigma^{n} &= \Sigma-\Sigma X^\top (X\Sigma X^\top+ T)^{-1} X \Sigma 
\end{align} \label{eq: gaussianposterior}
where $T$ is a $n$ by $n$ diagonal matrix with entries $\tau_i^2$, $1 \leq i \leq n$. An advantage of the Gaussian model is that it allows the posterior mean and covariance to be calculated directly without potentially computationally intensive MCMC algorithms. This makes the Gaussian model an easy to compute baseline to compare with more complicated Bayesian models. Furthermore, owing to the closed form of the posterior, we can bound the mean squared error (MSE) between the posterior mean $\bm{\mu}^{n}$ and the true parameter (stock changes and flows) values $\bm{\theta}^*$. A discussion on this is provided below, which gives some intuition why Bayesian methods can perform better than non Bayesian methods in the low data setting $n \ll p$.

While computationally convenient, an disadvantage with the Gaussian conjugate model is that the posterior distributions for flow variables can contain non negligible probability on negative values when there is insufficient data or uninformative priors, which is not physically meaningful as flow quantities should be strictly positive, making it not fully Bayesian in this sense. While the posterior mean of the Gaussian model can perform adequately in terms of point estimation, we do not recommend it for uncertainty quantification without very informative priors.


Nevertheless, the Gaussian model usefully gives some intuition into why Bayesian methods perform better than non Bayesian methods in the low data setting $n \ll p$. Below we provide some theoretical analysis for the mean squared error of the Gaussian model.

Owing to the closed form of the posterior of the Gaussian model described in the Methods section of the main text, we can bound the mean squared error (MSE) between the posterior mean $\bm{\mu}^{n}$ and the true parameter (stocks changes and flows) values $\bm{\theta}^*$ as follows:

\begin{theorem}\label{thm: GaussianMSEbound}

Let $\tau_i^2=\tau^2$, in other words $T=\tau^2 I$, $I$ the identity matrix. The mean squared error between the posterior mean of the Gaussian posterior $\bm{\mu}^{n}$ and the true stock change and flow values $\bm{\theta}^*$ can be bounded above in the following way:

\begin{align}
& \mathbb{E}(\| \bm{\theta}^*-\bm{\mu}^{n} \|^2_2)=\mathbb{E}(\sum_{j=1}^{p} (\theta_j^*-\mu_j^{n})^2) \nonumber \\
 & \leq \Tr(\Sigma^{-1/2}(\bm{\theta}^*-\bm{\mu})(\bm{\theta}^*-\bm{\mu})^\top \Sigma^{-1/2}) \lambda_1 (p-n + \sum_{j=1}^{n} \frac{\tau^4}{(d_j^2+\tau^2)^2}) + \tau^2 \sum_{j=1}^{n} \frac{\lambda_1 d_j^2}{(d_j^2+\tau^2)^2} \label{eq: gaussianmspe}
\end{align} 
where $\lambda_1\geq\lambda_2 \geq \dots \geq \lambda_p>0$ are the eigenvalues of $\Sigma$ in descending order, and $0 \leq d_1\leq d_2 \leq \dots \leq d_n $ are the singular values of $X\Sigma^{1/2}$ in ascending order. 

\end{theorem}

The bound in \Cref{eq: gaussianmspe} helps to give some insight into how the mean squared error of the model evolves as more data is added. This is perhaps easiest to see in the case when $\tau$ is near $0$ (meaning the data is almost noiseless) and $0 \leq \tau \ll d_1$ , which leaves 
$$\Tr(\Sigma^{-1/2}(\bm{\theta}^*-\bm{\mu})(\bm{\theta}^*-\bm{\mu})^\top \Sigma^{-1/2}) \lambda_1 (p-n)$$
as the dominant term. The $\Tr(\Sigma^{-1/2}(\bm{\theta}^*-\bm{\mu})(\bm{\theta}^*-\bm{\mu})^\top \Sigma^{-1/2})$ part of this term can be interpreted as a measurement of how close the prior mean is to the true values of the parameters (it is $0$ when $\bm{\theta}^*=\bm{\mu}$) and does not depend on the data, while the factor $\lambda_1 (p-n)$ decreases as $n$ increases, meaning as more data is added. This indicates that in a high dimensional setting $n \ll p$ which is inherent in MFA studies, the availability of informative priors is key to obtaining more accurate estimates of the true stock change and flow values. 

Furthermore, in the case when $\Sigma=\lambda_1 I$, \Cref{eq: gaussianmspe} can be directly compared with the ridge regression case. Since ridge regression is equivalent to the case when $\bm{\mu}=\bm{0}$ and $\Sigma=\lambda_1 I$, \Cref{eq: gaussianmspe} is smaller for the Gaussian model compared to ridge regression whenever $\Tr((\bm{\theta}^*-\bm{\mu})(\bm{\theta}^*-\bm{\mu})^\top)=\| \bm{\theta}^*-\bm{\mu} \|^2_2$ is smaller than $\|\bm{\theta}^*\|^2_2$. In other words whenever the prior mean $\bm{\mu}$ is closer to the true value in (L2) distance than the zero vector $\bm{0}$, which is consistent with the simulations presented in \Cref{sec: zinccasestudy}.

\begin{proof}[Proof of Theorem 1]

Recall that the posterior mean of the conjugate Gaussian model can be written as:

\begin{align*}
\bm{\mu}^{n}=\bm{\mu}+\Sigma X^\top (X\Sigma X^\top+\tau^2 I)^{-1} (X( \bm{\theta}^*-\bm{\mu})+\bm{\epsilon})
\end{align*}

This can be derived from looking at the joint distribution $[\bm{\theta}, 
X\bm{\theta}+\bm{\epsilon}]^\top$ and using standard Gaussian conditioning formulae:

\begin{align}
\begin{bmatrix} 
\bm{\theta} \\ 
X\bm{\theta}+\bm{\epsilon}
\end{bmatrix}
\sim \mathcal{N}\left(\begin{bmatrix} 
\bm{\mu} \\
X\bm{\mu}
\end{bmatrix},
\begin{bmatrix}
    \Sigma & \Sigma X^\top \\
   X \Sigma  & X \Sigma X^\top+\tau^2 I
   \end{bmatrix}\right)
\end{align}\label{eq: massconservationjointdist}
Suppose $dim(X)=(n,p)$, $dim(\Sigma)=(p,p)$, $dim(\bm{\mu})=dim(\bm{\theta}^*)=(p,1)$, $dim(\bm{\epsilon})=(n,1)$, and we'll assume $n<p$, since material flow analysis typically has less data than parameters. Assume $\Sigma$ is symmetric positive definite and $X$ has full rank. The mean squared error (MSE) between the posterior mean $\bm{\mu}^{n}$ and the true value $\bm{\theta}^*$ is equal to:

\begin{align*}
MSE&=\mathbb{E} \|\bm{\theta}^*-\bm{\mu}-\Sigma X^\top (X\Sigma X^\top+\tau^2 I)^{-1} (X( \bm{\theta}^*-\bm{\mu})+\bm{\epsilon})
  \|^2_2 \\
&=\mathbb{E} (\bm{\beta}-\Sigma X^\top (X\Sigma X^\top+\tau^2 I)^{-1} (X \bm{\beta}+\bm{\epsilon}))^\top (\bm{\beta}-\Sigma X^\top (X\Sigma X^\top+\tau^2 I)^{-1} (X \bm{\beta}+\bm{\epsilon})) \\
&=\mathbb{E} (\bm{\beta}^\top (I-\Sigma X^\top (X\Sigma X^\top+\tau^2 I)^{-1} X)^\top  (I-\Sigma X^\top (X\Sigma X^\top+\tau^2 I)^{-1} X) \bm{\beta}) \\
&+ \mathbb{E} (\bm{\epsilon}^\top (\Sigma X^\top (X\Sigma X^\top+\tau^2 I)^{-1})^\top  \Sigma X^\top (X\Sigma X^\top+\tau^2 I)^{-1} \bm{\epsilon})
\end{align*}
Where $\bm{\beta}=\bm{\theta}^*-\bm{\mu}$. At this point we decompose the MSE into the bias and variance terms. Let 
$$MSE_1=\mathbb{E} (\bm{\beta}^\top (I-\Sigma X^\top (X\Sigma X^\top+\tau^2 I)^{-1} X)^\top  (I-\Sigma X^\top (X\Sigma X^\top+\tau^2 I)^{-1} X) \bm{\beta})$$ denote the bias term.
and $$MSE_2=\mathbb{E} (\bm{\epsilon}^\top (\Sigma X^\top (X\Sigma X^\top+\tau^2 I)^{-1})^\top  \Sigma X^\top (X\Sigma X^\top+\tau^2 I)^{-1} \bm{\epsilon})$$ denote the variance term.

For the bias term, using the fact that the trace of a scalar is equal to itself and the cyclic property of trace, we have:

\begin{align*}
& MSE_1=\Tr(\bm{\beta}^\top (I-\Sigma X^\top (X\Sigma X^\top+\tau^2 I)^{-1} X)^\top  (I-\Sigma X^\top (X\Sigma X^\top+\tau^2 I)^{-1} X) \bm{\beta}) \\
&=\Tr(\bm{\beta}\bm{\beta}^\top (I-\Sigma X^\top (X\Sigma X^\top+\tau^2 I)^{-1} X)^\top  (I-\Sigma X^\top (X\Sigma X^\top+\tau^2 I)^{-1} X) ) \\
&=\Tr(\bm{\beta}\bm{\beta}^\top \Sigma^{-1/2}\Sigma^{1/2}(I-X^\top (X\Sigma X^\top+\tau^2 I)^{-1} X\Sigma ) \Sigma^{-1/2}\Sigma \Sigma^{-1/2}(I-\Sigma X^\top (X\Sigma X^\top+\tau^2 I)^{-1} X)  \Sigma^{1/2}\Sigma^{-1/2} ) \\
&=\Tr(\Sigma^{-1/2}\bm{\beta}\bm{\beta}^\top \Sigma^{-1/2}(I-\Sigma^{1/2} X^\top (X\Sigma X^\top+\tau^2 I)^{-1} X\Sigma^{1/2}) \Sigma (I-\Sigma^{1/2} X^\top (X\Sigma X^\top+\tau^2 I)^{-1} X\Sigma^{1/2})   )
\end{align*}
Notice $\Pi_{\tau}=\Sigma^{1/2} X^\top (X\Sigma X^\top+\tau^2 I)^{-1} X\Sigma^{1/2}$ is symmetric. Similarly $I-\Pi_{\tau}$ is symmetric and so $(I-\Pi_{\tau})^2$ is positive semidefinite. 

Let $X\Sigma^{1/2}=UDV^\top$ be the singular value decomposition of $X\Sigma^{1/2}$, where $d_1\leq d_2 \leq \dots \leq d_n$ are the diagonal elements of $D$ (also known as the singular values of $X\Sigma^{1/2}$) arranged in ascending order, then we have:

\begin{align}
\Pi_{\tau} &= VD^\top U^\top (UDD^\top U^\top+\tau^2 I)^{-1} UDV^\top = VD^\top (DD^\top +\tau^2 I)^{-1} DV^\top \nonumber
 \end{align}
 which has the same eigenvalues as $D^\top (DD^\top +\tau^2 I)^{-1} D$ by similarity (\cite{Horn1985}), which are $d_i^2/(d_i^2+\tau^2)$ and $0$ (repeated $p-n$ times). Likewise the eigenvalues of $I-\Pi_{\tau}$ are therefore $1$ (repeated $p-n$ times) and $\tau^2/(d_i^2+\tau^2)$

returning to the bias term, we have:

\begin{align}
MSE_1&=\Tr(\Sigma^{-1/2}\bm{\beta}\bm{\beta}^\top \Sigma^{-1/2}(I-\Pi_\tau) \Sigma (I-\Pi_\tau)) \nonumber \\ 
& \leq \Tr(\Sigma^{-1/2}\bm{\beta}\bm{\beta}^\top \Sigma^{-1/2}) \Tr((I-\Pi_\tau)^2 \Sigma ) \nonumber \\
& \leq \Tr(\Sigma^{-1/2}\bm{\beta}\bm{\beta}^\top \Sigma^{-1/2}) (\sum_{j=1}^{p-n} \lambda_j + \sum_{j=1}^{n} \lambda_{p-n+j} \tau^4/(d_j^2+\tau^2)^2) \nonumber \\
 & \leq  \Tr(\Sigma^{-1/2}\bm{\beta}\bm{\beta}^\top \Sigma^{-1/2}) \lambda_1 (p-n + \sum_{j=1}^{n} \tau^4/(d_j^2+\tau^2)^2)
 \end{align}
Where we used Von Neumann's trace inequality (\cite{Mirsky1975}) in the penultimate line, on the matrices $(I-\Pi_\tau)^2$ and $\Sigma$. So provided $\tau^2$ is small compared to the $d_j^2$, $MSE_1$ decreases approximately linearly in $n$, as $n$ increases to $p$.

For the variance term $MSE_2$, we once again use the singular value decomposition $X\Sigma^{1/2}=UDV^\top$:

\begin{align}
MSE_2&=\Tr(\mathbb{E} (\bm{\epsilon}^\top (\Sigma X^\top (X\Sigma X^\top+\tau^2 I)^{-1})^\top  \Sigma X^\top (X\Sigma X^\top+\tau^2 I)^{-1} \bm{\epsilon})) \nonumber  \\ 
&= \Tr(\mathbb{E} (\bm{\epsilon} \bm{\epsilon}^\top (\Sigma X^\top (X\Sigma X^\top+\tau^2 I)^{-1})^\top  \Sigma X^\top (X\Sigma X^\top+\tau^2 I)^{-1} )) \nonumber \\ 
&= \tau^2 \Tr( (X\Sigma X^\top+\tau^2 I)^{-2} X\Sigma^2 X^\top)  \nonumber
\\
& = \tau^2 \Tr(( UDD^\top U^\top+\tau^2 I)^{-2}UDV^\top \Sigma VD^\top U^\top) \nonumber \\
& = \tau^2 \Tr(U^\top (UDD^\top U^\top+\tau^2 I)^{-1} UU^\top (UDD^\top U^\top+\tau^2 I)^{-1}  UDV^\top \Sigma VD^\top ) \nonumber \\
& = \tau^2 \Tr( VD^\top (DD^\top +\tau^2 I)^{-2}  DV^\top \Sigma  ) \nonumber \\
& \leq \tau^2 \sum_{j=1}^{n} \lambda_1 d_j^2/(d_j^2+\tau^2)^2 \nonumber
\end{align}
Where the last line again follows from Von Neumann's trace inequality, on the matrices $VD^\top (DD^\top +\tau^2 I)^{-2}  DV^\top$ and $\Sigma$.

\end{proof}

\newpage

\subsection{Discussion of model parametrisation} \label{sec: parametrisationdiscussion}

In this section we discuss some additional limitations with the parametrisation of \cite{Gottschalk2010} which led us to use a mass based parametrisation. These limitations do not apply in all MFA systems but could potentially be problematic in some. In order to discuss them we provide a summary of the parametrisation: let $U_{i,j}$ denote the value of the flow from process $i$ to process $j$. Let $z_i=\sum_{k} U_{i,k}$ denote the total outflow of process $i$, and $\phi_{i,j}=U_{i,j}/z_i$ be the transfer coefficient for the flow from process $i$ to process $j$, equal to the fraction of the total outflow $z_i$ of process $i$. In addition, an optional external flow $q_i$ is allowed for each process $i$. Applying conservation of mass at process $i$, we obtain:

\begin{equation}
q_i+\sum_{j} U_{j,i} = \sum_{k} U_{i,k}
\end{equation}

which can be rewritten as 

\begin{equation} \label{eq: LuptonCoM} 
q_i+\sum_{j} z_j \phi_{j,i} = z_i
\end{equation} 

the conservation of mass equations \ref{eq: LuptonCoM} can therefore be written in the matrix form

\begin{equation} \label{eq:LuptonCoMmatrix} 
(I-\Phi^\top)\bm{z}=\bm{q}
\end{equation} 

where $\Phi$ is a matrix with elements $\phi_{i,j}$. \cite{Lupton2018} and \cite{Gottschalk2010} then notes that this equation can be inverted to give 
$(I-\Phi^\top)^{-1}\bm{q}=\bm{z}$, the vector of the total outflow of every process, which can then be used to retrieve each individual flow variables $U_{i,j}$ via the relationship $U_{i,j}=\phi_{i,j} z_i$. Using this, \cite{Lupton2018} and \cite{Gottschalk2010} assigns priors on the transfer coefficients $\phi_{i,j}$ and the external flow variables $q_i$, which gives an implicit prior over the flow variables $U_{i,j}$.

The first limitation is that it is unclear if $I-\Phi^\top$ is always invertible. From $\phi_{i,j}=U_{i,j}/\sum_{k} U_{i,k}$ we must have $\sum_{k} \phi_{i,k}=1$, which means the matrix $\Phi$ has each row summing to $1$ (making it a stochastic matrix), which implies it must have eigenvalue $1$ (this can be easily shown by checking the vector with entries all equal to $1$ is an eigenvector). $I-\Phi^\top$ therefore must have $0$ as an eigenvalue and is not invertible. 

It appears the way this issue is currently handled is to have at least one process in the system that has no outflows (see for example Figure 1 of \cite{Dong2022}, which has processes 6,7,8 and 9 without outflows), essentially fixing one or more rows of $\Phi$ to equal to the $\bm{0}$ vector so that it is no longer a stochastic matrix. However, it is unclear if a process that has no outflows always exists (for example Figure 2 of \cite{Mannan2020}) and such processes must be interpretable as not mass balanced (e.g. a stock or outflow of some kind). 



Additionally, if $\bm{q}$ is no longer nonnegative (e.g. if a stock change, external outflow, or mass balance relaxation term is added to the right hand side of \ref{eq:LuptonCoMmatrix} ) and assuming $I-\Phi^\top$ is invertible, it will cause the total flow output variables $\bm{z}=(I-\Phi^\top)^{-1}\bm{q}$ and by extension the flow variables $U_{i,j}$ to no longer be nonnegative. However, flow variables should be nonnegative to be physically meaningful. To see this, note that $(I-\Phi^\top)^{-1}$ has Taylor expansion $\sum_{m=0}^{\infty} (\Phi^\top)^m$, which is a limit of a sum of matrices with nonnegative entries, so the entries of $(I-\Phi^\top)^{-1}$ must all be nonnegative. Therefore in general $\bm{z}=(I-\Phi^\top)^{-1}\bm{q}$ is nonnegative if and only if $\bm{q}$ is nonnegative. This leaves how to incorporate mass balance relaxations (in order to take into account epistemic uncertainty) and other nonnegative terms under \cite{Gottschalk2010}'s parametrisation a possible direction for future work. 



\end{document}